\documentclass[12pt]{article}%
\usepackage{amsmath}
\usepackage{graphicx}%
\usepackage{amsfonts}%
\usepackage{amssymb}
\usepackage{color}
\usepackage{bm}
\usepackage{natbib}
\usepackage{authblk}
\usepackage{comment}
\usepackage{chngcntr}

\newcommand{\bet}{ \mbox{\boldmath $ \eta $} }

\newcommand{\bbeta}{ \mbox{\boldmath $ \beta $} }

\newcommand{\balpha}{ \mbox{\boldmath $ \alpha $} }

\newcommand{\btheta}{ \mbox{\boldmath $ \theta $} }

\newcommand{\bzero}{\textbf{0}}

\newcommand{\bh}{\textbf{h}}

\newcommand{\bI}{\textbf{I}}

\newcommand{\bs}{\textbf{s}}

\newcommand{\bV}{\textbf{V}}

\newcommand{\bW}{\textbf{W}}

\newcommand{\bX}{\textbf{X}}

\title{Preferential Sampling for Bivariate Spatial Data}
\author[1]{Shinichiro Shirota\thanks{Corresponding author. Email address : shinichiro.shirota@gmail.com}}
\author[2]{Alan E. Gelfand}
\affil[1]{Department of Commerce, Meiji University, 1-1, Kanda-Surugadai, Chiyoda-ku, Tokyo 101-8301, Japan}
\affil[1]{Center for Social Data Science Education and Research Promotion, Hitotsubashi University, 2-1 Naka, Kunitachi-shi, Tokyo 186-8601, Japan}
\affil[2]{Department of Statistical Science, Duke University, 2080 Duke University Road, Durham, NC 27708, USA}

\begin{document}
\maketitle

\vspace*{0.5cm}
\begin{abstract}

Preferential sampling provides a formal modeling specification to capture the effect of bias in a set of sampling locations on inference when a geostatistical  model is used to explain observed responses at the sampled locations.  In particular, it enables modification of spatial prediction adjusted for the bias.  Its original presentation in the literature addressed assessment of the presence of such sampling bias while follow on work focused on regression specification to improve spatial interpolation under such bias.  All of the work in the literature to date considers the case of a univariate response variable at each location, either continuous or modeled through a latent continuous variable.  The contribution here is to extend the notion of preferential sampling to the case of bivariate response at each location.  This exposes sampling scenarios where both responses are observed at a given location as well as scenarios where, for some locations, only one of the responses is recorded.  That is, there may be different sampling bias for one response than for the other.  It leads to assessing the impact of such bias on co-kriging.  It also exposes the possibility that preferential sampling can bias inference regarding dependence between responses at a location.  We develop the idea of bivariate preferential sampling through various model specifications and illustrate the effect of these specifications on prediction and dependence behavior.  We do this both through simulation examples as well as with a forestry dataset that provides mean diameter at breast height (MDBH) and trees per hectare (TPH) as the point-referenced bivariate responses.

\end{abstract}


\section{Introduction}

Since the concept was introduced into the literature by \cite{Diggleetal(10)}, preferential sampling (PS) has attracted considerable attention, resulting in a literature we briefly review below.  For a region of interest, the basic issue is bias in the sampling of spatial locations where point referenced response data are collected and the potential impact on inference for the response surface over the region.  The canonical illustrative example addresses the objective of inferring about environmental exposures.  If environmental monitors are only placed in locations where environmental levels tend to be high, then interpolation based upon observations from these stations will necessarily produce only high predictions.  The obvious remedy lies in suitable spatial design of the locations.  For example, a random or space-filling design \citep[][]{SaltzmanNychka(98)} for locations over the region of interest is expected to preclude such bias.

However, sampling  may not be designed in this fashion.  Environmental researchers may install monitoring stations where they expect to find high exposure levels; ecologists may tend to sample where they expect to find individuals.  Recognizing the possibility of such bias, can prediction be revised to adjust for it?  Specifically, the intention of PS modeling is to address two questions. First, is there evidence of a PS effect?  Second, can we improve prediction in the presence of PS?  We do not seek to remove PS; we do not propose to revise the data collection.  Rather, we seek to acknowledge its presence and attempt to mitigate its impact.

Our contribution here is to expand the issue of sampling bias to the case of bivariate response, in order to address the effect on co-kriging.  Recall that, if responses at location $\bs$, say $(Y_1(\bs), Y_2(\bs))$ are dependent then prediction of say $Y_1(\bs_0)$ at an unobserved location $\bs_{0}$ can benefit from using all of the observed $Y_2$'s in addition to all of the observed $Y_1$'s \citep{Wackernagel(03), BanerjeeCarlinGelfand(14)}.  Then, the question becomes the following.  Suppose $Y_{1}(\bs)$ and $Y_{2}(\bs)$ are dependent spatial observations.  That is, for a given region $D$, realizations of the response surfaces, $Y_{1}(\bs): \bs \in D$ and $Y_{2}(\bs): \bs \in D$ exhibit strong correlation\footnote{We formalize what this means in our modeling specifications below.}.  Then, if there is sampling bias in the subsets of locations, $\mathcal{S}_1$ and $\mathcal{S}_{2}$ in $D$ where $Y_1$ and $Y_2$ are sampled, respectively, can we formalize and demonstrate the presence of a \emph{bivariate} PS effect?  Can we improve co-kriging through a model which captures bivariate PS?  Further, can we assess whether bivariate PS impacts the local dependence between the responses?

In the ensuing development we offer simulation illustration and also consider an illustrative data example. In particular, for the real illustration, we use forest inventory data collected in the Penobscot Experimental
Forest (PEF), Maine, under a unit of the Northern Research Station, U.S. Forest Service.
Point-referenced data is observed at 589 forest inventory plots
across PEF.  At a site we use mean tree diameter at breast height (MDBH) (breast height is measured at 1.37m above the forest floor) and tree density measured as number of trees per hectare (TPH) as our responses.  Here, the ``constant yield law'' \citep{WeinerFreckleton(10)} argues that, at equilibrium, the total yield/biomass on a plot is roughly constant regardless of the number of individuals, i.e., size of individuals will decrease as density increases, suggesting strong negative dependence.
We consider prediction of MDBH and TPH as continuous surfaces across the study region in Section 3, then investigate correlation between MDBH and TPH under PS in Section 4.

Avoiding formal details for the moment, we recall that PS, as defined by \cite{Diggleetal(10)}, considers  $\mathcal{Y}$, the response data collected and $\mathcal{S}$, the set of locations where the data was collected as two random objects.  The first is a partial realization of the response process $Y(\bs)$ over $D$; the second is a realization of a point pattern over $D$.  Then, PS arises if $\mathcal{Y}$ and $\mathcal{S}$ are stochastically dependent (they are obviously functionally dependent since $Y(\bs_i) \in \mathcal{Y}$ is observed at $\bs_i \in \mathcal{S}$).

In the bivariate setting we have $\mathcal{Y}_1$ with associated $\mathcal{S}_1$ and $\mathcal{Y}_2$ with associated $\mathcal{S}_{2}$.  This raises two possibilities: (i) $\mathcal{S}_{1} = \mathcal{S}_{2}$ and (ii) $\mathcal{S}_{1} \neq \mathcal{S}_{2}$.  For instance, ozone and temperature data responses could fall under (i) if they are collected at the same set of monitoring stations.  We imagine a single point pattern but bivariate geostatistical response.  Is there evidence of PS for $\mathcal{Y}_{1}$? For $\mathcal{Y}_{2}$? Does PS influence one response differently from the other. We treat this setting in Section 3.1.

For possibility (ii), of practical interest, we consider the case of $\mathcal{S}_{1} \neq \mathcal{S}_{2}$ but not disjoint.  We view this as a missing data situation, i.e., $\mathcal{S} = \mathcal{S}_1 \cup \mathcal{S}_2 $ with $\mathcal{Y}_1$ observed on $\mathcal{S}_1$ and $\mathcal{Y}_2$ observed on $\mathcal{S}_{2}$.  Additionally, depending upon the nature of the data collection, e.g., the first response is more difficult to record, we can have $\mathcal{S}_1 \subset \mathcal{S}_2$.  In any event, we can have differential sampling bias for the two responses.  The MDBH and TPH can fall under this case (ii) because it may be fairly efficient to count the total number of tree but more demanding to measure the diameters of each tree.  We treat this setting in Section 3.2, asking similar questions to those in Sections 3.1.

Possibility (ii) can also arise when the responses were collected for different analyses, perhaps with different protocols so we can have $\mathcal{S}_1 \cap \mathcal{S}_2 = \phi$.  Data of this sort, dependent spatial variables collected at location sets that are disjoint but over the same region, are not found in the literature.  However, we do note that, with dependence between the latent Gaussian processes that drive the geostatistical modeling for $Y_{1}(\bs)$ and $Y_{2}(\bs)$, prediction for say $Y_{1}(\bs_{0})$ can benefit from both $\mathcal{Y}_1$ and $\mathcal{Y}_2$, even if  $\mathcal{S}_1$ and $\mathcal{S}_2$ are disjoint.  We clarify this in Appendix A.  So, in Section 3.3, using simulation, we add PS for $\mathcal{Y}_1$ and for $\mathcal{Y}_2$ to reveal whether we can learn about PS in the disjoint point pattern case as well as whether we can improve on geostatistical prediction introducing PS.

A novel opportunity that the bivariate setting brings is to examine how PS can affect inference regarding the dependence between $Y_{1}(\bs)$ and $Y_{2}(\bs)$ and more generally, between $Y_{1}(\bs)$ and $Y_{2}(\bs')$, the cross-covariance behavior. That is, perhaps not surprisingly, sampling bias for bivariate spatial data can affect inference on the dependence structure, i.e., second order behavior, in addition to first order inference.  We take this up in Section 4.


Apart from the forestry dataset we employ, other potential settings include ozone ($O_3$), and fine particulate matter ($PM_{2.5}$) data with different monitoring networks for the different responses providing different point patterns; selling prices of different types of residential properties with point patterns arising as locations of the sales \citep{Pacietal(20)};  crime data with point patterns for different crime types and response being say time of day for the crime \citep{ShirotaGelfand(17a)}.
Further, we can imagine modifying the geostatistical model to spatial generalized linear models \citep{DiggleTawnMoyeed(98)}.  For example, we might consider binary response to study presence/absence for say a pair of species across sites.  Using joint species distribution models (e.g., \cite{Thorsonetal(15), Ovaskainenetal(16)}, we obtain a shared point pattern for say two species. See \cite{GelfandShirota(19)} in this regard. Alternatively, the response may be abundance, say perhaps, count data, basal area or percent ground cover.  Regardless, we can ask whether PS is different for one species vs. the other.  Further, we can introduce environmental covariates so that some apply to the point patterns while others apply to the geostatistical specifications.

In Section 2 we briefly review the formalities of and customary models for preferential sampling with a univariate response.  In Section 3 we develop the bivariate response setting, considering three different sampling scenarios, each with its own preferential sampling interpretation.  In Section 4 we offer a brief excursion into the potential effect of preferential sampling on local bivariate dependence.  Section 5 concludes with a brief summary and possible future work.

\section{Brief Review of Preferential Sampling}

The notion of PS was introduced into the literature in the seminal paper of \cite{Diggleetal(10)}.  Relevant follow-on modeling papers in this regard are \cite{Patietal(11)} and \cite{GelfandShirota(19)}. PS assumes that the set of sampling locations is a realization of a spatial point process though it may not have been developed randomly. That is, it may be designed in some fashion and be deterministic but, in practice, not necessarily with the intention of being roughly uniformly distributed over $D$.  Then, as noted in the Introduction, the question becomes a stochastic one:  is the realization of the responses independent of the realization of locations?  If not, then we have what is called PS.  The dependence here is stochastic dependence.  Notationally/functionally, the responses are associated with the locations.

Consider the general notation $\boldsymbol{\eta}$, ${\cal S}$ and ${\cal Y}$ to denote a latent process, the design locations, and the measurement data, respectively.  Using bracket notation for density functions, a general factorization of the joint
distribution of $\boldsymbol{\eta}$, ${\cal S}$ and ${\cal Y}$ is
$[\boldsymbol{\eta},{\cal S},{\cal Y}] = [\boldsymbol{\eta}][{\cal S}|\boldsymbol{\eta}][{\cal Y}|{\cal S},\boldsymbol{\eta}]$.
This factorization is most natural from a modelling perspective because
(i) the latent process drives both data mechanisms and (ii) ${\cal S}$ and ${\cal Y}$ can be viewed as a marked point process \citep{Illianetal(08)} where, in the geostatistical setting, we would model locations and then response given location.

Under non-PS, $[{\cal S}|\boldsymbol{\eta}]=[{\cal S}]$ so the stochastic variation in ${\cal S}$ can be ignored for inference about $\boldsymbol{\eta}$ or ${\cal Y}$.
Conventional geostatistical methods
do this, treating the design as a fixed set of locations $\bs_i: i=1,...,n$
and, typically, assuming the measurements $Y(\bs_i)$ are conditionally independent given
the corresponding $\eta(\bs_i)$.  Hence, $[{\cal Y}|{\cal S},\boldsymbol{\eta}] = \prod_{i=1}^n [Y(\bs_i)|\eta(\bs_i)]$.

Under a geostatistical model for $Y(\bs)$ of the form $Y(\bs) = \mu(\bs) +\eta(\bs) +\epsilon(\bs)$ with $\eta(\bs)$ a Gaussian process (GP),
\cite{Diggleetal(10)}
assume that ${\cal S}$ is a log Gaussian Cox Poisson process (LGCP)
with intensity $\lambda(\bs) = \exp\{\alpha + \gamma \eta(\bs)\}$.
The common $\eta(\bs)$ yields what is referred to as a ``shared process'' model.
Non-PS arises when $\gamma=0$, and, otherwise, we have strong PS.

\cite{Patietal(11)} propose expanded modeling, adding a second Gaussian process.
Again, assume that ${\cal S}$ is a LGCP, now writing the  intensity as $\lambda(\bs) = \exp\{\alpha + \eta(\bs)\}$
where $\eta(\bs)$ is a mean zero Gaussian process with say, covariance function
$\sigma_{1}^{2} \rho(\bh/\phi_1)$ for separation vector $\bh$.
The measurements $Y(\bs_i): i=1,...,n$ at locations
$\bs_i$ follow the model $Y(\bs) = \mu(\bs) + \gamma \text{log} \lambda(\bs) + w(\bs) + \epsilon(\bs)$
where $w(\bs)$ is a mean zero Gaussian process, independent of $\eta(\bs)$, with
mean zero with covariance function $\sigma_{2}^{2} \rho(\bh/\phi_2)$.
The $\epsilon(\bs)$ are pure error $N(0,\tau^2)$ variables; we restore the general geostatistical model when $\gamma=0$.  Note that the $\gamma$ has been moved from the LGCP model in \cite{Diggleetal(10)} to serve as a regression coefficient in \cite{Patietal(11)}.
We have a re-parametrized version of \cite{Diggleetal(10)} when the process $w(\bs)$ is absent, i.e., $\sigma_2^2=0$.
The parameter $\gamma$ controls the degree of \emph{preferentiality} in the sampling of the $Y(\bs)$.
The process $w(\bs)$ allows for a component of the spatial variation in the response process that is not linked to the
sampling process.

\subsection{Univariate PS modeling}

We briefly summarize some competing models for the univariate PS case.
Specifically, we model ${\cal S}$ using a LGCP driven by an intensity $\lambda(\bs)$ specified as
\begin{equation}
\log\lambda(\bs) = \bX(\bs)^{T}\bbeta + \eta(\bs),
\label{eq:loglambda}
\end{equation}
where $\bX(\bs)$ is a $p$-dimensional covariate vector at location $\bs$.  The $\eta(\bs)$ are spatial random effects which come from a zero mean GP with covariance function denoted by $C(\bs, \bs^\prime) = \sigma_{\eta}^{2}\rho_{\eta}(\bs - \bs^\prime; \theta_\eta)$, where $\rho_{\eta}(\cdot; \theta_\eta)$ is a correlation function depending on $\theta_\eta$.
The LGCP likelihood for a realization $\mathcal{S}$ is given by
\begin{equation}
\mathcal{L}_1\left(\bbeta, \bet_{D}, \sigma_\eta^2;\, \mathcal{S}\right)  =\exp\left\{-\int_D{\lambda(\bs)d\bs}\right\} \prod_{i=1}^n \lambda(\bs_i),
\label{eq:LGCPlike}
\end{equation}
where $\bet{_{D}}= \{\eta(\bs): \bs \in D\}$.  The integral in (\ref{eq:LGCPlike}) is stochastic, i.e., an integral over a random realization of a stochastic process.  It can never be evaluated explicitly and, in practice, it is approximated numerically using a grid of \emph{representative points} \citep{BanerjeeCarlinGelfand(14)} within $D$.

Turning to the responses, let $\bW(\bs) = \left(\bV(\bs),\bX(\bs)\right)$ where $\bV(\bs)$ is a $q$-dimensional vector, i.e., an augmented covariate vector to associate with $Y(\bs)$.
We consider four regression models for prediction of $Y(\bs)$:

(i)  A simple spatial regression model
\begin{equation}
Y(\bs) = \bW(\bs)^{T}\balpha + \varepsilon(\bs),
\label{eq:regr}
\end{equation}
where we partition $\balpha = (\balpha_{V},\balpha_{X})$.  The $\varepsilon(\bs)$ are white noise errors, normally distributed with zero mean and variance $\tau^{2}$. This model is fitted apart from the point pattern model, excluding the opportunity to learn about ${\cal Y}$ from ${\cal S}$ through the intensity.

(ii) A shared component model

\begin{equation}
Y(\bs) = \bW(\bs)^{T}\balpha + \delta\eta(\bs) + \varepsilon(\bs),
\label{eq:shared}
\end{equation}
where $\eta(\bs)$ is as in (\ref{eq:loglambda}).  Again, $\balpha = (\balpha_{V},\balpha_{X})$ and the $\varepsilon(\bs)$ are white noise errors, normally distributed with zero mean and variance $\tau^{2}$.  Here, $\eta(\bs)$ plays the role of a regressor and $\delta$ becomes the coefficient for a PS effect with the sign of $\delta$ indicating the direction of preferential adjustment.  So, fitting (4) and (1) jointly enables assessment of the \emph{presence} of PS.

(iii) A geostatistical model
\begin{equation}
Y(\bs) = \bW(\bs)^{T}\balpha + \phi(\bs) + \varepsilon(\bs),
\label{eq:spatialGP}
\end{equation}
where again, $\balpha = (\balpha_{V},\balpha_{X})$ and the $\varepsilon(\bs)$ are white noise errors, normally distributed with zero mean and variance $\tau^{2}$.  Here,  the $\phi(\bs)$ are spatial random effects specified by a zero mean GP with covariance function of the form $C(\bs - \bs^\prime; \theta_\phi) = \sigma_{\phi}^{2}\rho_{\phi}(\bs-\bs^\prime; \theta_\phi)$, where $\rho_\phi (\cdot; \theta_\phi)$ is a correlation function depending on $\theta_\phi$.  They provide local adjustment to the model in \eqref{eq:regr}.
This model is also fitted apart from the point pattern model, excluding the opportunity to learn about ${\cal Y}$ from ${\cal S}$ through the intensity.

(iv) A geostatistical model with PS
\begin{equation}
Y(\bs) = \bW(\bs)^{T}\balpha + \phi(\bs) + \delta \eta(\bs) + \varepsilon(\bs),
\label{eq:twoGP}
\end{equation}
where $\eta(\bs)$ is as in (\ref{eq:loglambda}) and, again, $\balpha = (\balpha_{V},\balpha_{X})$. The $\phi(\bs)$ are spatial random effects as in (\ref{eq:spatialGP}), and the $\varepsilon(\bs)$ are white noise errors, normally distributed with zero mean and variance $\tau^{2}$.  Again, we have a shared process form and the coefficient $\delta$ carries the interpretation of a PS effect. The $\eta$ process and the $\phi$ process are modeled, a priori, as independent.  We see the introduction of $\eta(\bs)$ as a regressor in the geostatistical model.
Because of the flexibility of $\phi(\bs)$,  model (iv) need not perform better than model (iii).  So, joint fitting of (6) and (1) assesses whether $\delta$ remains significant, whether we can \emph{improve} on geostatistical prediction of $Y$ in the presence of PS.

\section{Modeling, inference, and prediction under bivariate preferential sampling}

We now formalize the notion of preferential sampling in the bivariate response case.  Letting $\mathcal{S}_{1}$ denote the point pattern of locations associated with the first response and $\mathcal{S}_{2}$ the point pattern of locations associated with the second response, we consider three scenarios.  The first, in Section 3.1, assumes $\mathcal{S}_{1} = \mathcal{S}_{2}$, that is, a common set of sampling sites is associated with the two responses.  The second, in Section 3.2, assumes $\mathcal{S}_{1} \neq \mathcal{S}_{2}$ but that they are not disjoint.  Here, we imagine a missing data setting where, say, at a given location, observation of one of the responses may be difficult to record.  Lastly, in Section 3.3 we assume $\mathcal{S}_{1} \cap \mathcal{S}_{2} = \phi$, that is, each response variable is associated with its own point pattern and the generative models for the two point patterns are not the same. We can imagine two different data collection efforts but over the same region.  Each subsection is considered with both a simulation example and a real data example (using the data described in Section 3.1.2).

\subsection{$\mathcal{S}_{1} = \mathcal{S}_{2}$}

We first consider the case where the observations at a location arise in pairs.  Here, $\mathcal{S}_{1} = \mathcal{S}_{2} = \mathcal{S}$ and, with ${\cal S} = \{\bs_1, \bs_2,..., \bs_{n}\}$, we have ${\cal Y}_{1} = \{Y_1(\bs_1), Y_1(\bs_2),...,Y_1(\bs_{n})\}$ and  ${\cal Y}_{2} = \{Y_2(\bs_1), Y_2(\bs_2),...,Y_2(\bs_{n})\}$.  As in Section 2, we specify a LGCP for ${\cal S}$ with intensity log$\lambda(\bs) = \bX^{T}(\bs)\balpha +  \eta(\bs)$ where $\eta(\bs)$ is a GP as above.  Simplifying the covariates to just $\bX(\bs)$, we envision four specifications for the $Y$'s that are analogues of the models of the previous section:

(i) $M_{1}^{=}$: $Y_1(\bs) = \bX^{T}(\bs) \bbeta_{1} + \epsilon_1(\bs)$ and
$Y_2(\bs) = \bX^{T}(\bs) \bbeta_{2} + \epsilon_2(\bs)$.

(ii) $M_{2}^{=}$: $Y_1(\bs) = \bX^{T}(\bs) \bbeta_{1} + \gamma_{1}\eta(\bs)  + \epsilon_1(\bs)$ and
$Y_2(\bs) = \bX^{T}(\bs) \bbeta_{2} + \gamma_{2}\eta(\bs) + \epsilon_2(\bs)$.

(iii) $M_{3}^{=}$: $Y_1(\bs) = \bX^{T}(\bs) \bbeta_{1} + a_{11}\omega_{1}(\bs)   + \epsilon_1(\bs)$ and
$Y_2(\bs) = \bX^{T}(\bs) \bbeta_{2} + a_{21}\omega_{1}(\bs) + a_{22}\omega_{2}(\bs) + \epsilon_2(\bs)$.

(iv) $M_{4}^{=}$: $Y_1(\bs) = \bX^{T}(\bs) \bbeta_{1} + \gamma_{1}\eta(\bs) +a_{11}\omega_{1}(\bs)   + \epsilon_1(\bs)$ and
$Y_2(\bs) = \bX^{T}(\bs) \bbeta_{2} + \gamma_{2}\eta(\bs) + a_{21}\omega_{1}(\bs) + a_{22}\omega_{2}(\bs) + \epsilon_2(\bs)$.
As we proceed from model $M_{1}^{=}$ to model $M_{4}^{=}$, we introduce $0$, $1$, $2$, and $3$ GPs, respectively.

Here, $M_{1}^{=}$ provides spatial regressions for $Y_{1}(\bs)$ and $Y_{2}(\bs)$ with no spatial random effects.
$M_{2}^{=}$ is a shared process model with the process for ${\cal S}$ being inserted as a regressor to explain each of the $Y$'s. The shared process makes the two responses spatially dependent with readily calculated dependence structure. Specifically, the cross-covariance matrix is $c_{\eta}(\bs - \bs')\left(
                             \begin{array}{cc}
                               \gamma_{1}^{2} & \gamma_{1}\gamma_{2} \\
                               \gamma_{1}\gamma_{2} & \gamma_{2}^{2} \\
                             \end{array}
                           \right)$.
$M_{3}^{=}$ is a customary coregionalized geostatistical model, using a lower triangular specification, which makes the responses spatially dependent but ignores PS. The cross covariance matrix here is $\rho_{\omega_{1}}(\bs-\bs')\left(
                                                                                    \begin{array}{cc}
                                                                                      a_{11}^{2} & a_{11}a_{21} \\
                                                                                      a_{11}a_{21} & a_{21}^{2} \\
                                                                                    \end{array}
                                                                                  \right)$ + $\rho_{\omega_{2}}(\bs-\bs')\left(
                                                                                    \begin{array}{cc}
                                                                                      0 & 0 \\
                                                                                      0 & a_{22}^{2} \\
                                                                                    \end{array}
                                                                                  \right)$.
So, $M_{2}^{=}$ and $M_{3}^{=}$ specify different types of dependence for the $Y$'s. $M_{4}^{=}$ includes both types of spatial dependence with cross covariance matrix the sum of those for $M_2^{=}$ and $M_{3}^{=}$.

The primary questions here are the following:  $M_{2}^{=}$ focuses on whether we can show a significant $\gamma$ coefficient for either or both responses? This model mimics the original \cite{Diggleetal(10)} setting, trying to find a PS story.   Also, we can ask whether the coefficients are different from each other?  A noteworthy limitation for this model is the fact that, with only a single GP, both geostatistical responses have the same range.  
With two GP's, do we expect $M_{3}^{=}$ to outperform $M_{2}^{=}$?  The flexibility of the coregionalization, not having both responses share a common GP, suggests that this will be the case.  However, the model comparison can show that PS explains the responses nearly as well.
In this regard, we can set $a_{22}=0$, a dimension reduction to a single process which also implies that both geostatistical responses have the same range, the so-called \emph{separable} specification \citep{BanerjeeCarlinGelfand(14)}.   Lastly, we can compare the first three models against $M_{4}^{=}$.   Under $M_{4}^{=}$, can we still find significant $\gamma$'s in the presence of coregionalization?  This is analogous to the analysis in \cite{Patietal(11)} where they were able to find significant PS in the presence of the usual geostatistical model.

For both simulated and real data, the modeling and data analysis is implemented in all cases in a Bayesian setting using weak priors and Markov chain Monte Carlo (MCMC) to fit the model.
For the posterior samples of Gaussian processes ($\bm{\eta}$ and $\bm{w}$ surfaces) and their hyperparameters, we implement elliptical slice sampling for the GPs and hyperparameters following \cite{MurrayAdamsMacKay(10), MurrayAdams(10)}.
For the other parameters, random walk Metropolis-Hastings and Gibbs sampling are implemented.
Model comparison is done through cross-validation since, with an LGCP specification for the point pattern, validation through hold out of a random subset of the data is well-known \citep{GelfandSchliep(18)}.

Specifically, we use $50\%$ of the data for the training set and $50\%$ for the testing set. We construct the training sets in two different ways. First, we hold out $50\%$ of the locations at random, fitting the remaining $50\%$.  Second, we fit the $50\%$ of the locations with the largest response values.  With bivariate response, we can implement this bias to the fitting set using either of the response variables.   In fact, we do this separately for each response and develop the predictive performance assessment for the responses individually and jointly.  Predictive performance criteria used are the predictive root mean square error (RMSE) and the continuous ranked probability score  \citep[CRPS,][]{GneitingRaftery(07)}.  In fact, we compute these for each response separately as well as in total.  With simulated data where we introduce sampling bias in the generative model, we expect the PS story to emerge with random holdout and become more exaggerated with further bias in the fitting data.  With real data, depending upon the actual data collection design employed, we may not see strong PS; adding further sampling bias to the fitting data can help to illuminate the story.  Within the PS setting the sole model assessment objective is predictive performance with regard to the responses, so we do not consider assessment of the point pattern model.  However, if we fit $\mathcal{Y}$ and $\mathcal{S}$ jointly, then $\mathcal{Y}$ does inform about the LGCP for $\mathcal{S}$.

\subsubsection{Simulation examples}

Following the discussion above, we demonstrate parameter recovery and present model comparison based on predictive performance for a simulated data illustration.
The data is simulated from model $M_{2}^{=}$ assuming  $\log \lambda(\bs) =\bX^{\top}(\bs)\balpha+\eta(\bs)$ with $\bX=(1, s_{y})^{\top}$ and $\eta(\bs) \sim \text{GP}(0, C(||\bs-\bs'||; \btheta)$.
       Further, $Y_{1}(\bs) =\bX^{\top}(\bs)\bbeta_{1}+\gamma_{1}\eta(\bs)+\epsilon_{1}(\bs)$ with $\epsilon_{1}(\bs)\sim \mathcal{N}(0, \tau_{1}^2)$ and
        $Y_{2}(\bs) =\bX^{\top}(\bs)\bbeta_{2}+\gamma_{2}\eta(\bs)+\epsilon_{2}(\bs)$ with $\epsilon_{2}(\bs)\sim \mathcal{N}(0, \tau_{2}^2)$.


We assume a unit square region $D=[0, 1]\times [0, 1]$, and employ a $30\times 30 = 900$ regular grid for cells over the region to generate a sufficiently high resolution realization from $\eta(\bs)$ as the representative points to evaluate the stochastic integrals for the  LGCP likelihood. Here, $\bX(\bs)$ contains an intercept along with $s_{y}$ the $y$ coordinate minus mean (0.5) as the regressors. We adopt an exponential covariance for $\eta(\bs)$, i.e., $\sigma^2 \exp(-\phi \|\bs-\bs'\|)$.
We consider both a small ($\sigma^2=1/3$) and a large ($\sigma^2=1$) variance for the $\eta(\bs)$ surface which, again, is realized at the centroids of the $900$ grid cells. The point pattern $\mathcal{S}$ is simulated from $\text{LGCP}(\lambda(\cdot))$ by the Poisson thinning approach \citep{LewisShedler(79)}.  Then, we simulate response surfaces on $\mathcal{S}$ choosing $\eta(\bs)$ evaluated at the representative point of the nearest grid.
The response data is simulated under the parameter values: $(\alpha_1, \alpha_2) = (6, 1), \quad \sigma^2 = 1 (\text{or}  1/3), \quad \phi = 3,   (\beta_{11}, \beta_{21}) = (0, 0.5), \quad (\beta_{12}, \beta_{12}) = (0, 0.5), \quad (\gamma_{1}, \gamma_{2}) = (1, 0.3) \quad (\tau_{1}^2, \tau_{2}^2) = (0.3, 0.1).$

%

That is, the preferential sampling effects are specified as $\gamma_{1}=1$, a relatively large value and $\gamma_{2}=0.3$, a relatively small value. The total number of locations in $\mathcal{S}$ is 487 for the low variance case, 554 for the high variance case.  Weakly informative priors for all parameters are adopted as: $\balpha \sim \mathcal{N}(\bzero, 100  \bI_{p}), \quad \sigma^2 \sim \mathcal{IG}(2, 0.1), \quad \phi \sim \mathcal{U}(0, 100), \bbeta_{1}, \bbeta_{2} \sim \mathcal{N}(\bzero, 100  \bI_{p}), \quad
        \gamma_{1}, \gamma_{2} \sim \mathcal{N}(0, 100), \quad \tau_{1}^2, \tau_{2}^2 \sim \mathcal{IG}(2, 0.1).$


In the fitting, we discard the first 10,000 iterations as burn-in and preserve the subsequent 20,000 as posterior samples. The MCMC iterations
are tuned so that effective sample sizes (ESS) for all parameters are larger than 200.
As above, we choose the following holdout strategies: (I) the random order, (II-a) the descending order of the first response, and (II-b) the descending order of the second response.
For (I), we preserve randomly 50$\%$ of the locations for testing and fit the model to the remaining 50$\%$ locations. For (II-a) and (II-b), we select $100p\%$ of the locations using the descending order of each response (with (II-a) for the first response and (II-b) for the second response).  Then, we select $100(0.5-p)\%$ of the locations randomly to complete the set of training locations. The remaining 50$\%$ locations are preserved as the set of test locations. We set $p=0.2$ and $p=0.35$ to control the degree of preferential sampling effect (a larger $p$ introduces a stronger PS effect). This strategy strengthens the PS effect although, according to the model specification, the random holdout itself includes a PS effect.

Table \ref{tab:sim_pred_S1=S2} shows the model comparison of predictive performance for the simulated data for low ($\sigma^2 \phi=1$) and high ($\sigma^2 \phi=3$) variance. It reveals that $M_{2}^{=}$ and $M_{4}^{=}$ have similar performance with respect to CRPSs and RMSEs for both cases and better performance than $M_{3}^{=}$.
Again, our data is generated under $M_2^{=}$ but in Table \ref{tab:sim_est_S1=S2} of Appendix C we present parameter recovery results for all four models above. Altogether, the true values of parameters are recovered well in $M_{2}^{=}$; notably $\gamma_{1}, \gamma_{2}$ are also recovered in $M_{4}^{=}$ though $M_{4}^{=}$ adds Gaussian processes. The estimated values of the identifiable parameters for Gaussian processes \citep{Zhang(04)}, $a_{1, 1}^2\phi_{w1}$, $a_{2, 2}^2\phi_{w2}$, are very small. The correlation parameter $a_{2,1}$ is also insignificant, suggesting that $M_{4}^{=}$ distinguishes the shared process from the coregionalization. Figure \ref{fig:sim_int_S1=S2} in Appendix B shows the estimated log intensity surfaces for the low and the high variance cases. These surfaces are recovered well by both $M_{2}^{=}$ and $M_{4}^{=}$.

This  simulation exercise is primarily illustrative, one of many that we have explored.
The takeaway points are as follows.  If the true model is, in fact, a PS model like $M_2^{=}$, we can recover the PS story.  With small and large variance in the shared process and with no bias in the sampling, we find $M_2^{=}$ and $M_4^{=}$ are preferred though the difference compared with $M_3^{=}$ is small.  Further, when we introduce sampling bias in the fitting data, we also find  $M_2^{=}$ and $M_4^{=}$ are preferred.  Additionally, the stronger $\gamma$ for the first response reveals larger associated RMSE's and CPRS's.

\begin{table}[htbp]
\caption{Model comparison of predictive performance for the simulated data in the $\mathcal{S}_{1}=\mathcal{S}_{2}$ case for low ($\sigma^2=1/3$) and high ($\sigma^2=1$) variance with $p=0.2$ and $p=0.35$ overlap rate. The bold fonts indicate the best performance for each holdout}
\begin{center}
\scalebox{0.80}{
\begin{tabular}{lcccccccccccccc}
\hline
\hline
low/$p=0.2$ &  \multicolumn{4}{c}{(I)} & & \multicolumn{4}{c}{(II-a)} & & \multicolumn{4}{c}{(II-b)} \\
 & $M_{1}^{=}$ & $M_{2}^{=}$  & $M_{3}^{=}$ & $M_{4}^{=}$ & & $M_{1}^{=}$ & $M_{2}^{=}$ & $M_{3}^{=}$ & $M_{4}^{=}$ & & $M_{1}^{=}$ & $M_{2}^{=}$ & $M_{3}^{=}$ & $M_{4}^{=}$ \\
\hline
 $\text{RMSE}_{1}$  & 0.706 & 0.605 & 0.613 & 0.611 & & 0.722 & 0.586 & 0.606 & 0.583 & & 0.685 & 0.613 & 0.605 & 0.623 \\
 $\text{RMSE}_{2}$  & 0.342 & 0.333 & 0.334 & 0.334 & & 0.336 & 0.316 & 0.318 & 0.317 & & 0.362 & 0.320 & 0.339 & 0.317 \\
  $\text{RMSE}_{1+2}$ & 1.048 & $\bm{0.938}$ & 0.947 & 0.945 & & 1.058 & 0.902 & 0.924 & $\bm{0.900}$ & & 1.047 & $\bm{0.933}$ & 0.944 & 0.940 \\
 \hline
$\text{CRPS}_{1}$ & 0.540 & 0.402 & 0.412 & 0.410 & & 0.526 & 0.371 & 0.395 & 0.382 & & 0.515 & 0.368 & 0.385 & 0.372 \\
 $\text{CRPS}_{2}$ & 0.251 & 0.224 & 0.228 & 0.227 & & 0.252 & 0.214 & 0.219 & 0.220 & & 0.266 & 0.208 & 0.230 & 0.205 \\
 $\text{CRPS}_{1+2}$  & 0.791 & $\bm{0.626}$ & 0.640 & 0.637 & & 0.778 & $\bm{0.585}$ & 0.614 & 0.602 & & 0.781 & $\bm{0.576}$ & 0.615 & 0.577 \\
\hline
\hline
low/$p=0.35$ &  \multicolumn{4}{c}{(I)} & & \multicolumn{4}{c}{(II-a)} & & \multicolumn{4}{c}{(II-b)} \\
 & $M_{1}^{=}$ & $M_{2}^{=}$  & $M_{3}^{=}$ & $M_{4}^{=}$ & & $M_{1}^{=}$ & $M_{2}^{=}$ & $M_{3}^{=}$ & $M_{4}^{=}$ & & $M_{1}^{=}$ & $M_{2}^{=}$ & $M_{3}^{=}$ & $M_{4}^{=}$ \\
\hline
 $\text{RMSE}_{1}$  & - & - & - & - & & 0.932 & 0.690 & 0.749 & 0.684 & & 0.735 & 0.619 & 0.614 & 0.624 \\
 $\text{RMSE}_{2}$  & - & - & - & - & & 0.379 & 0.342 & 0.351 & 0.335 & & 0.442 & 0.393 & 0.410 & 0.398 \\
  $\text{RMSE}_{1+2}$ & - & - & - & - & & 1.311 & 1.032 & 1.100 & $\bm{1.019}$ & & 1.177 & $\bm{1.012}$ & 1.024 & 1.022 \\
 \hline
$\text{CRPS}_{1}$ & - & - & - & - & & 0.761 & 0.491 & 0.534 & 0.477 & & 0.550 & 0.392 & 0.400 & 0.391 \\
 $\text{CRPS}_{2}$ & - & - & - & - & & 0.285 & 0.242 & 0.250 & 0.234 & & 0.357 & 0.282 & 0.304 & 0.285 \\
 $\text{CRPS}_{1+2}$ & - & - & - & - & & 1.046 & 0.733 & 0.784 & $\bm{0.711}$ & & 0.907 & $\bm{0.674}$ & 0.704 & 0.676 \\
\hline
\hline
high/$p=0.2$ &  \multicolumn{4}{c}{(I)} & & \multicolumn{4}{c}{(II-a)} & & \multicolumn{4}{c}{(II-b)} \\
 & $M_{1}^{=}$ & $M_{2}^{=}$  & $M_{3}^{=}$ & $M_{4}^{=}$ & & $M_{1}^{=}$ & $M_{2}^{=}$ & $M_{3}^{=}$ & $M_{4}^{=}$ & & $M_{1}^{=}$ & $M_{2}^{=}$ & $M_{3}^{=}$ & $M_{4}^{=}$ \\
\hline
 $\text{RMSE}_{1}$ & 0.895 & 0.690 & 0.683 & 0.676 & & 0.959 & 0.663 & 0.725 & 0.664 & & 0.988 & 0.706 & 0.709 & 0.708 \\
 $\text{RMSE}_{2}$ & 0.365 & 0.320 & 0.326 & 0.321 & & 0.382 & 0.323 & 0.328 & 0.324 & & 0.395 & 0.301 & 0.323 & 0.298 \\
 $\text{RMSE}_{1+2}$ & 1.260 & 1.010 & 1.009 & $\bm{0.997}$ & & 1.341 & $\bm{0.986}$ & 1.053 & 0.988 & & 1.383 & 1.007 & 1.032 & $\bm{1.006}$ \\
 \hline
$\text{CRPS}_{1}$ & 0.663 & 0.440 & 0.442 & 0.416 & & 0.698 & 0.408 & 0.472 & 0.397 & & 0.741 & 0.433 & 0.460 & 0.433 \\
 $\text{CRPS}_{2}$ & 0.266 & 0.208 & 0.216 & 0.203 & & 0.284 & 0.208 & 0.216 & 0.205 & & 0.294 & 0.192 & 0.215 & 0.190 \\
 $\text{CRPS}_{1+2}$  & 0.929 & 0.648 & 0.658 & $\bm{0.619}$ & & 0.982 & 0.616 & 0.688 & $\bm{0.603}$ & & 1.035 & 0.625 & 0.675 & $\bm{0.623}$ \\
\hline
\hline
high/$p=0.35$ &  \multicolumn{4}{c}{(I)} & & \multicolumn{4}{c}{(II-a)} & & \multicolumn{4}{c}{(II-b)} \\
 & $M_{1}^{=}$ & $M_{2}^{=}$  & $M_{3}^{=}$ & $M_{4}^{=}$ & & $M_{1}^{=}$ & $M_{2}^{=}$ & $M_{3}^{=}$ & $M_{4}^{=}$ & & $M_{1}^{=}$ & $M_{2}^{=}$ & $M_{3}^{=}$ & $M_{4}^{=}$ \\
\hline
 $\text{RMSE}_{1}$ & - & - & - & - & & 1.275 & 0.744 & 0.939 & 0.730 & & 0.896 & 0.715 & 0.714 & 0.725 \\
 $\text{RMSE}_{2}$ & - & - & - & - & & 0.388 & 0.301 & 0.320 & 0.305 & & 0.479 & 0.383 & 0.422 & 0.380 \\
 $\text{RMSE}_{1+2}$ & - & - & - & - & & 1.663 & 1.045 & 1.259 & $\bm{1.035}$ & & 1.375 & $\bm{1.098}$ & 1.136 & 1.105 \\
 \hline
$\text{CRPS}_{1}$ & - & - & - & - & & 1.048 & 0.470 & 0.664 & 0.454 & & 0.659 & 0.427 & 0.464 & 0.437 \\
 $\text{CRPS}_{2}$ & - & - & - & - & & 0.292 & 0.195 & 0.208 & 0.195 & & 0.390 & 0.265 & 0.314 & 0.262 \\
 $\text{CRPS}_{1+2}$  & - & - & - & - & & 1.340 & 0.665 & 0.872 & $\bm{0.649}$ & & 1.049 & $\bm{0.692}$ & 0.778 & 0.699 \\
\hline
\hline
\end{tabular}
}
\end{center}
\label{tab:sim_pred_S1=S2}
\end{table}

\subsubsection{MDBH-TPH data}

As noted in the Introduction, here and through the remainder of the paper, we analyze forest inventory data collected in the Penobscot Experimental Forest (PEF), Maine, under a unit of the Northern Research Station, U.S. Forest Service.
This point-referenced data is observed at 589 forest inventory plots
across PEF. \cite{Sendaketal(03)} describe the PEF sampling design and the
outcome variables measured at each inventory plot location.
We use only mean tree diameter at breast height (MDBH) (breast height is measured at 1.37m above the forest floor) and tree density measured as number of trees per hectare (TPH). These variables are commonly used, in combination with other information, to assess the economic and ecological value of a forest. In the analysis below, each MDBH and TPH is transformed to the log scale.
The relationship between MDBH and TPH is influenced by individual and environmental factors. Environmental regressors include quality of soil, quantity of water and light, and competition for these resources. Forest disturbance history (e.g., timber harvesting, fire, or windthrow) also strongly influences MDBH and TPH. These factors often vary spatially and at different scales across a forest and are too expensive or impossible to measure directly.  As noted in the Introduction, in accord with the constant yield law, it is generally expected that, for an established forest, as MDBH increases the number of trees per hectare decreases.

We use remotely sensed predictor variables, recorded at a high spatial resolution across the PEF, to improve prediction at unobserved locations.
These predictor variables are derived from NASA’s Laser Vegetation Imaging Sensor (LVIS;
https://lvis.gsfc.nasa.gov) airborne waveform Light Detection and Ranging (LiDAR) sensor.
These LiDAR signals are high dimensional and highly correlated.  To avoid multicollinearity issues, a singular value decomposition (SVD) is employed to extract orthogonal vectors that explain some portion of the variance in LiDAR signals (see \cite{FinleyBanerjee(13)} for more details). The first four singular vectors, which explain over 90$\%$ of the signal variance, are used here. The interpolated surfaces for these covariates are displayed in Figure \ref{fig:MT_cov}.

\begin{figure}[ht]
  \begin{center}
   \includegraphics[width=15cm]{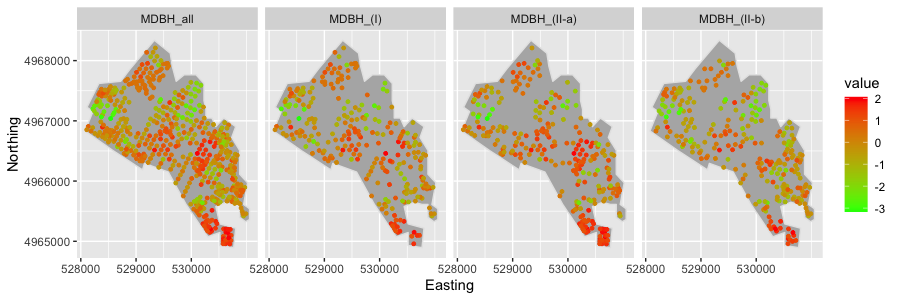}
  \end{center}
  \begin{center}
   \includegraphics[width=15cm]{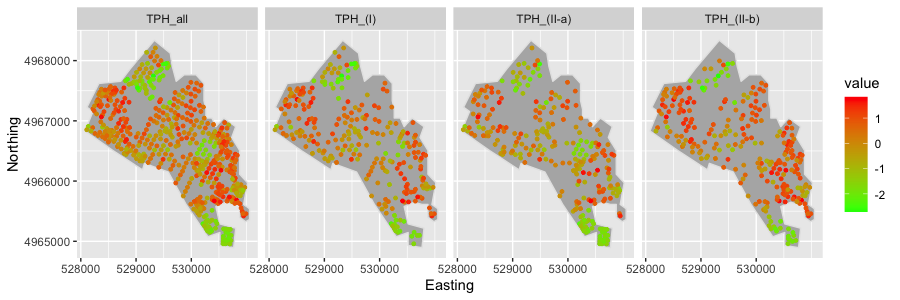}
  \end{center}
  \caption{The plots of MDBH (top) and TPH (bottom) on all locations (left) and locations preserved by each holdout strategy for $\mathcal{S}_{1}=\mathcal{S}_{2}$. (I):random, (II-a):partially overlapped with the descending order of MDBH, (II-b):partially overlapped with the descending order of TPH}
  \label{fig:MT_Y_S1=S2}
\end{figure}

The overall goal of the analysis is to predict MDBH and TPH as continuous
surfaces across the PEF. Our intention here is to learn about a potential bivariate preferential sampling story with regard to these responses under common sampling locations $\mathcal{S}=\mathcal{S}_{1}=\mathcal{S}_{2}$ for MDBH and TPH. We design $\mathcal{S}_{1}= \mathcal{S}_{2}$ from the 589 locations using the two strategies adopted in the previous subsection.
Figure \ref{fig:MT_Y_S1=S2} shows the plots of MDBH and TPH over the 589 locations and these values on the training set of locations selected by each strategy. The correlation between MDBH and TPH for all locations is -0.76, consistent with the constant yield law. (We investigate the effect of preferential sampling on correlation in Section 4.)

For inference and model comparison, we specify the LGCP likelihood for $\mathcal{S}=\mathcal{S}_{1}=\mathcal{S}_{2}$ as above. We implement regular grid approximation to the stochastic integral over the region, with a total of 732 regular grid cells. The eastings and northings are standardized so that the maximum distance is equal to 1, similar to the scale in simulation studies.
The same prior settings and MCMC fitting as in Section 3.1.1 are employed.
We provide the details of the estimation results in Table \ref{tab:MT_est_S1=S2_all}. The estimated $\gamma$'s by $M_{2}^{=}$ and $M_{4}^{=}$ have opposite signs, consistent with the negative correlation between MDBH and TPH over the observed locations. Furthermore, these gammas are significant for $M_{4}^{=}$. $a_{2,1}$ in $M_{3}^{=}$ is also negative, again supporting the constant yield law. Interestingly, even with significant opposite signs for gammas, $a_{2,1}$ is significantly negative for $M_{4}^{=}$.  This finding suggests that the negative correlation between MDBH and TPH might be driven by both preferential effects and the responses themselves.

Table \ref{tab:MT_pred_S1=S2} shows the predictive performance results. In summary, $M_2^{=}$ shows better performance than the other models with respect to CRPS. On the other hand, $M_4^{=}$ shows better (or nearly better) performance than the other models with respect to RMSE, $M_3^{=}$ also shows similar performance and outperforms $M_2^{=}$ with this measure. Altogether, $M_4^{=}$ shows good, competitive results with both measures.
Interestingly, even with the hold out strategy (I), $M_2^{=}$ and $M_4^{=}$ outperform $M_3^{=}$ with CRPS. This result implies the existence of preferential within the collected data itself.

\begin{table}[htbp]
\caption{The model comparison of predictive performance for the MDBH-TPH data in the $\mathcal{S}_{1}= \mathcal{S}_{2}$ case with $p=0.2$ and $p=0.35$ overlap rate.
The bold fonts suggest the best performance for each random and descending holdout}
\begin{center}
\scalebox{0.80}{
\begin{tabular}{lcccccccccccccc}
\hline
\hline
$p=0.2$ &  \multicolumn{4}{c}{(I)} & & \multicolumn{4}{c}{(II-a)} & & \multicolumn{4}{c}{(II-b)} \\
 & $M_{1}^{=}$ & $M_{2}^{=}$  & $M_{3}^{=}$ & $M_{4}^{=}$ & & $M_{1}^{=}$ & $M_{2}^{=}$ & $M_{3}^{=}$ & $M_{4}^{=}$ & & $M_{1}^{=}$ & $M_{2}^{=}$ & $M_{3}^{=}$ & $M_{4}^{=}$ \\
\hline
RMSE &  &  &  &  & &  &  &  &  & &  &  &  &  \\
MDBH & 0.954 & 0.755 & 0.690 & 0.685 & & 0.945 & 0.744 & 0.726 & 0.714 & & 0.917 & 0.838 & 0.754 & 0.760 \\
 TPH & 0.980 & 0.660 & 0.605 & 0.605 & & 1.058 & 0.702 & 0.620 & 0.614 & & 1.035 & 0.845 & 0.733 & 0.728 \\
 MDBD+TPH & 1.934 & 1.415 & 1.295 & $\bm{1.290}$ & & 2.003 & 1.446 & 1.346 & $\bm{1.328}$ & & 1.952 & 1.683 & $\bm{1.487}$ & 1.488 \\
 \hline
CRPS &  &  &  &  & &  &  &  &  & &  &  &  &  \\
MDBH & 0.695 & 0.455 & 0.475 & 0.461 & & 0.673 & 0.443 & 0.505 & 0.482 & & 0.679 & 0.527 & 0.531 & 0.523 \\
TPH & 0.736 & 0.378 & 0.423 & 0.414 & & 0.837 & 0.410 & 0.447 & 0.414 & & 0.705 & 0.502 & 0.522 & 0.501 \\
MDBH+TPH & 1.431 & $\bm{0.833}$ & 0.898 & 0.875 & & 1.510 & $\bm{0.853}$ & 0.952 & 0.896 & & 1.384 & 1.029 & 1.053 & $\bm{1.024}$ \\
\hline
\hline
$p=0.35$ &  \multicolumn{4}{c}{(I)} & & \multicolumn{4}{c}{(II-a)} & & \multicolumn{4}{c}{(II-b)} \\
 & $M_{1}^{=}$ & $M_{2}^{=}$  & $M_{3}^{=}$ & $M_{4}^{=}$ & & $M_{1}^{=}$ & $M_{2}^{=}$ & $M_{3}^{=}$ & $M_{4}^{=}$ & & $M_{1}^{=}$ & $M_{2}^{=}$ & $M_{3}^{=}$ & $M_{4}^{=}$ \\
\hline
RMSE &  &  &  &  & &  &  &  &  & &  &  &  &  \\
MDBH & - & - & - & - & & 1.290 & 0.966 & 0.996 & 0.967 & & 1.171 & 0.824 & 0.866 & 0.826 \\
 TPH & - & - & - & - & & 1.271 & 0.863 & 0.843 & 0.822 & & 1.416 & 0.795 & 0.847 & 0.798 \\
 MDBD+TPH & - & - & - & - & & 2.561 & 1.829 & 1.839 & $\bm{1.789}$ & & 2.587 & $\bm{1.619}$ & 1.713 & 1.624 \\
 \hline
CRPS &  &  &  &  & &  &  &  &  & &  &  &  &  \\
MDBH & - & - & - & - & & 1.003 & 0.650 & 0.730 & 0.680 & & 0.967 & 0.533 & 0.601 & 0.562 \\
TPH & - & - & - & - & & 1.090 & 0.564 & 0.642 & 0.579 & & 1.088 & 0.493 & 0.611 & 0.551 \\
MDBH+TPH & - & - & - & - & & 2.093 & $\bm{1.214}$ & 1.372 & 1.259 & & 2.055 & $\bm{1.026}$ & 1.212 & 1.113 \\
\hline
\hline
\end{tabular}
}
\end{center}
\label{tab:MT_pred_S1=S2}
\end{table}

\subsection{$\mathcal{S}_{1} \neq \mathcal{S}_{2}$ but not disjoint}

We view $\mathcal{S}_{1} \neq \mathcal{S}_{2}$ but not disjoint as a missing data case.  We can still imagine that $M_2$ generates the data, i.e., under a point pattern $\mathcal{S}$ and then $\mathcal{Y}_{1}$ and $\mathcal{Y}_{2}$.  This makes it clear that there is only one $\eta(\bs)$ driving the point pattern $\mathcal{S}$ yielding the $\bs$ points with
$\mathcal{Y}_{1}$ observed for $\mathcal{S}_{1} \subset \mathcal{S}$ and $\mathcal{S}_{2} \subset \mathcal{S}$.

\subsubsection{Simulation examples}

We investigate parameter recovery and predictive performance in $\mathcal{S}_{1} \neq \mathcal{S}_{2}$ but overlapping case.
We use the same datasets with low and high variance as in section 3.1.1, but separate the datasets into training and test sets by the two strategies described in Section 3.1.1.  To avoid strong overlap in the datasets we keep 50$\%$ of all locations as the training data, and preserve the remaining 50$\%$ of locations as the test data.
The first version (I) randomly selects $\mathcal{S}_{1}$ and $\mathcal{S}_{2}$ each as $100p\%$ commonly and $100(0.5-p)\%$ independently from all locations. The second version (II) proceeds as follows. First, as the set of training locations, we select $100p\%$ of all locations by the descending order of each response (denoted by (II-a) for the first response and (II-b) for the second response) and randomly select $100(0.5-p)\%$  for each response from the remaining set of locations. So, $p\%$ of the locations are common for both responses but the remaining set of fitting locations is different for each response. Again, $p$ controls the degree of preferential sampling introduced.  The larger $p$ is, the stronger the PS effects become for (II). For (I), $p$ controls the degree of overlapping but has no influence on PS effects.
For fitting and inference, we specify the LGCP likelihood for $\mathcal{S}=\mathcal{S}_{1}\cup \mathcal{S}_{2}$. The same prior settings and MCMC iterations as in Section 3.1.1 are employed.

Table \ref{tab:sim_pred_S1neqS2} shows the model comparison of predictive performance for the simulated data for low ($\sigma^2 \phi=1$) and high ($\sigma^2 \phi=3$) variance. For (I), $M_{2}^{\neq}$, $M_{3}^{\neq}$ and $M_{4}^{\neq}$ present similar performance; this is plausible since the Gaussian processes in $M_{3}^{\neq}$ capture the $\eta(\bs)$ surface when the locations are selected randomly.
For (II), the table shows that $M_{2}^{\neq}$ and $M_{4}^{\neq}$ have similar and better performance with respect to CRPS and RMSE than the other models for the low and high variance cases.
However, $M_{2}^{\neq}$ reveals slightly better performance than $M_{4}^{\neq}$ under stronger preferential sampling effects and larger overlapping, e.g., high/$p=0.35$ with (II-a).  We show the estimation results for the random holdout case (I) in Table \ref{tab:sim_est_S1neqS2} in Appendix C.
The $\gamma$'s are recovered well, as in the $\mathcal{S}_{1} = \mathcal{S}_{2}$ case.

\begin{table}[htbp]
\caption{Model comparison of predictive performance for the simulated data in the $\mathcal{S}_{1}\neq \mathcal{S}_{2}$ case for low ($\sigma^2=1/3$) and high ($\sigma^2=1$) variance with $p=0.2$ and $p=0.35$ overlap rate. The bold fonts suggest the best performance for each holdout}
\begin{center}
\scalebox{0.80}{
\begin{tabular}{lcccccccccccccc}
\hline
\hline
low/$p=0.2$ &  \multicolumn{4}{c}{(I)} & & \multicolumn{4}{c}{(II-a)} & & \multicolumn{4}{c}{(II-b)} \\
 & $M_{1}^{\neq}$ & $M_{2}^{\neq}$  & $M_{3}^{\neq}$ & $M_{4}^{\neq}$ & & $M_{1}^{\neq}$ & $M_{2}^{\neq}$ & $M_{3}^{\neq}$ & $M_{4}^{\neq}$ & & $M_{1}^{\neq}$ & $M_{2}^{\neq}$ & $M_{3}^{\neq}$ & $M_{4}^{\neq}$ \\
\hline
 $\text{RMSE}_{1}$  & 0.737 & 0.616 & 0.622 & 0.619 & & 0.759 & 0.576 & 0.612 & 0.586 & & 0.685 & 0.591 & 0.582 & 0.596 \\
 $\text{RMSE}_{2}$  & 0.336 & 0.316 & 0.320 & 0.317 & & 0.366 & 0.331 & 0.340 & 0.331 & & 0.385 & 0.348 & 0.371 & 0.348 \\
  $\text{RMSE}_{1+2}$ & 1.073 & $\bm{0.932}$ & 0.942 & 0.936 & & 1.125 & $\bm{0.907}$ & 0.952 & 0.917 & & 1.070 & $\bm{0.939}$ & 0.953 & 0.944 \\
 \hline
$\text{CRPS}_{1}$ & 0.553 & 0.418 & 0.428 & 0.417 & & 0.566 & 0.362 & 0.392 & 0.363 & & 0.517 & 0.360 & 0.372 & 0.367 \\
 $\text{CRPS}_{2}$ & 0.245 & 0.216 & 0.223 & 0.216 & & 0.277 & 0.237 & 0.244 & 0.232 & & 0.283 & 0.228 & 0.252 & 0.228 \\
 $\text{CRPS}_{1+2}$  & 0.798 & 0.634 & 0.651 & $\bm{0.633}$ & & 0.843 & 0.599 & 0.636 & $\bm{0.595}$ & & 0.800 & $\bm{0.588}$ & 0.624 & 0.595 \\
\hline
\hline
low/$p=0.35$ &  \multicolumn{4}{c}{(I)} & & \multicolumn{4}{c}{(II-a)} & & \multicolumn{4}{c}{(II-b)} \\
 & $M_{1}^{\neq}$ & $M_{2}^{\neq}$  & $M_{3}^{\neq}$ & $M_{4}^{\neq}$ & & $M_{1}^{\neq}$ & $M_{2}^{\neq}$ & $M_{3}^{\neq}$ & $M_{4}^{\neq}$ & & $M_{1}^{\neq}$ & $M_{2}^{\neq}$ & $M_{3}^{\neq}$ & $M_{4}^{\neq}$ \\
\hline
 $\text{RMSE}_{1}$  & 0.709 & 0.604 & 0.612 & 0.606 & & 0.923 & 0.671 & 0.731 & 0.695 & & 0.743 & 0.633 & 0.623 & 0.621 \\
 $\text{RMSE}_{2}$  & 0.339 & 0.328 & 0.328 & 0.329 & & 0.376 & 0.345 & 0.347 & 0.342 & & 0.426 & 0.387 & 0.407 & 0.387 \\
  $\text{RMSE}_{1+2}$ & 1.048 & $\bm{0.932}$ & 0.940 & 0.935 & & 1.299 & $\bm{1.016}$ & 1.078 & 1.037 & & 1.169 & 1.020 & 1.030 & $\bm{1.008}$ \\
 \hline
$\text{CRPS}_{1}$ & 0.550 & 0.401 & 0.401 & 0.399 & & 0.755 & 0.461 & 0.524 & 0.492 & & 0.559 & 0.407 & 0.400 & 0.396 \\
 $\text{CRPS}_{2}$ & 0.251 & 0.226 & 0.227 & 0.225 & & 0.281 & 0.236 & 0.242 & 0.238 & & 0.338 & 0.284 & 0.301 & 0.283 \\
 $\text{CRPS}_{1+2}$  & 0.801 & 0.627 & 0.628 & $\bm{0.624}$ & & 1.036 & $\bm{0.697}$ & 0.766 & 0.730 & & 0.897 & 0.691 & 0.701 & $\bm{0.679}$ \\
\hline
\hline
high/$p=0.2$ &  \multicolumn{4}{c}{(I)} & & \multicolumn{4}{c}{(II-a)} & & \multicolumn{4}{c}{(II-b)} \\
 & $M_{1}^{\neq}$ & $M_{2}^{\neq}$  & $M_{3}^{\neq}$ & $M_{4}^{\neq}$ & & $M_{1}^{\neq}$ & $M_{2}^{\neq}$ & $M_{3}^{\neq}$ & $M_{4}^{\neq}$ & & $M_{1}^{\neq}$ & $M_{2}^{\neq}$ & $M_{3}^{\neq}$ & $M_{4}^{\neq}$ \\
\hline
 $\text{RMSE}_{1}$ & 0.934 & 0.682 & 0.692 & 0.695 & & 0.926 & 0.654 & 0.744 & 0.685 & & 0.988 & 0.712 & 0.727 & 0.720 \\
 $\text{RMSE}_{2}$ & 0.340 & 0.291 & 0.287 & 0.289 & & 0.366 & 0.294 & 0.307 & 0.302 & & 0.376 & 0.297 & 0.317 & 0.299 \\
 $\text{RMSE}_{1+2}$ & 1.274 & $\bm{0.973}$ & 0.979 & 0.984 & & 1.292 & $\bm{0.948}$ & 1.051 & 0.987 & & 1.364 & $\bm{1.009}$ & 1.044 & 1.019 \\
 \hline
$\text{CRPS}_{1}$ & 0.714 & 0.417 & 0.439 & 0.419 & & 0.690 & 0.388 & 0.477 & 0.422 & & 0.742 & 0.432 & 0.459 & 0.442 \\
 $\text{CRPS}_{2}$ & 0.256 & 0.192 & 0.197 & 0.188 & & 0.276 & 0.186 & 0.207 & 0.197 & & 0.284 & 0.189 & 0.207 & 0.191 \\
 $\text{CRPS}_{1+2}$  & 0.970 & 0.609 & 0.636 & $\bm{0.607}$ & & 0.966 & $\bm{0.574}$ & 0.684 & 0.619 & & 1.026 & $\bm{0.621}$ & 0.666 & 0.633 \\
\hline
\hline
high/$p=0.35$ &  \multicolumn{4}{c}{(I)} & & \multicolumn{4}{c}{(II-a)} & & \multicolumn{4}{c}{(II-b)} \\
 & $M_{1}^{\neq}$ & $M_{2}^{\neq}$  & $M_{3}^{\neq}$ & $M_{4}^{\neq}$ & & $M_{1}^{\neq}$ & $M_{2}^{\neq}$ & $M_{3}^{\neq}$ & $M_{4}^{\neq}$ & & $M_{1}^{\neq}$ & $M_{2}^{\neq}$ & $M_{3}^{\neq}$ & $M_{4}^{\neq}$ \\
\hline
 $\text{RMSE}_{1}$ & 0.881 & 0.711 & 0.695 & 0.715 & & 1.177 & 0.672 & 0.845 & 0.695 & & 0.900 & 0.691 & 0.682 & 0.693 \\
 $\text{RMSE}_{2}$ & 0.362 & 0.323 & 0.320 & 0.323 & & 0.379 & 0.301 & 0.307 & 0.303 & & 0.439 & 0.343 & 0.377 & 0.340 \\
 $\text{RMSE}_{1+2}$ & 1.243 & 1.034 & $\bm{1.015}$ & 1.038 & & 1.556 & $\bm{0.973}$ & 1.152 & 0.998 & & 1.339 & 1.034 & 1.059 & $\bm{1.033}$ \\
 \hline
$\text{CRPS}_{1}$ & 0.638 & 0.441 & 0.443 & 0.451 & & 0.993 & 0.429 & 0.586 & 0.452 & & 0.675 & 0.416 & 0.426 & 0.415 \\
 $\text{CRPS}_{2}$ & 0.264 & 0.206 & 0.212 & 0.207 & & 0.282 & 0.195 & 0.197 & 0.196 & & 0.359 & 0.236 & 0.271 & 0.232 \\
 $\text{CRPS}_{1+2}$  & 0.902 & $\bm{0.647}$ & 0.655 & 0.658 & & 1.275 & $\bm{0.624}$ & 0.783 & 0.648 & & 1.034 & 0.652 & 0.697 & $\bm{0.647}$ \\
\hline
\hline
\end{tabular}
}
\end{center}
\label{tab:sim_pred_S1neqS2}
\end{table}

\subsubsection{MDBH, TPH data}

Now, we investigate $\mathcal{S}_{1}\neq \mathcal{S}_{2}$ case with the MDBH-TPH data used in Section 3.1.2. We create distinct sampling locations $\mathcal{S}_{1}$ and $\mathcal{S}_{2}$ for MDBH and TPH, respectively. We design $\mathcal{S}_{1}\neq \mathcal{S}_{2}$ from the 589 locations using the two strategies adopted in the previous subsection.
Figure \ref{fig:MT_Y_S1neqS2} shows the plots of MDBH and TPH on the 589 locations and these values on the training set of locations selected by each strategy. 

For inference and model comparison, we specify the LGCP likelihood for $\mathcal{S}=\mathcal{S}_{1}\cup \mathcal{S}_{2}$ as above. Again, we implement regular grid approximation to the stochastic integrals over the region, with the total number of grid cells taken as 732. The same prior settings and MCMC iterations as in Section 3.1 are employed.
Instead of estimation results at all locations (this case is available only for $\mathcal{S}_{1}=\mathcal{S}_{2}$), we include the details of the estimation results under case (I) with $p=0.2$ in Table \ref{tab:MT_est_S1neqS2_random}. Case (I) does not introduce artificial sampling bias but can capture preferential sampling effects the original data might have.

The estimated $\gamma$'s by $M_{2}^{\neq}$ have opposite signs, again consistent with the negative correlation between MDBH and TPH on observed locations. Coefficient $a_{2,1}$ in $M_{3}^{\neq}$ is also negative, again supporting the constant yield law. For $M_{4}^{\neq}$, the estimated $\gamma$'s are not significant but $a_{2,1}$ is significantly negative.

Table \ref{tab:MT_pred_S1neqS2} shows the predictive performance results.
For (I), $M_3^{\neq}$ and $M_4^{\neq}$ reveal similar performance with respect to RMSE, but, for CRPS, $M_2^{\neq}$ shows better performance than the other models with overlapping rate $p=0.35$. For (II), $M_3^{\neq}$ and $M_4^{\neq}$ show similar better performance with respect to RMSE, but $M_2^{\neq}$ shows better performance with CRPS than other models. $M_4^{\neq}$ shows better CRPS performance than $M_3^{\neq}$, so $M_4^{\neq}$ is a competitive model with both RMSE and CRPS.

\begin{table}[htbp]
\caption{The model comparison of predictive performance for the MDBH-TPH data in the $\mathcal{S}_{1}\neq \mathcal{S}_{2}$ case with $p=0.2$ and $p=0.35$ overlap rate.
The bold fonts suggest the best performance for each random and descending holdout}
\begin{center}
\scalebox{0.80}{
\begin{tabular}{lcccccccccccccc}
\hline
\hline
$p=0.2$ &  \multicolumn{4}{c}{(I)} & & \multicolumn{4}{c}{(II-a)} & & \multicolumn{4}{c}{(II-b)} \\
 & $M_{1}^{\neq}$ & $M_{2}^{\neq}$  & $M_{3}^{\neq}$ & $M_{4}^{\neq}$ & & $M_{1}^{\neq}$ & $M_{2}^{\neq}$ & $M_{3}^{\neq}$ & $M_{4}^{\neq}$ & & $M_{1}^{\neq}$ & $M_{2}^{\neq}$ & $M_{3}^{\neq}$ & $M_{4}^{\neq}$ \\
\hline
RMSE &  &  &  &  & &  &  &  &  & &  &  &  &  \\
MDBH & 0.978 & 0.805 & 0.727 & 0.724 & & 0.969 & 0.771 & 0.759 & 0.763 & & 0.990 & 0.826 & 0.750 & 0.739 \\
 TPH & 1.007 & 0.719 & 0.657 & 0.656 & & 1.129 & 0.803 & 0.691 & 0.694 & & 1.115 & 0.718 & 0.686 & 0.684 \\
 MDBD+TPH & 1.985 & 1.524 & 1.384 & $\bm{1.380}$ & & 2.098 & 1.574 & $\bm{1.450}$ & 1.457 & & 2.105 & 1.544 & 1.436 & $\bm{1.423}$ \\
 \hline
CRPS &  &  &  &  & &  &  &  &  & &  &  &  &  \\
MDBH & 0.733 & 0.520 & 0.492 & 0.480 & & 0.695 & 0.488 & 0.527 & 0.520 & & 0.743 & 0.497 & 0.497 & 0.484 \\
TPH & 0.763 & 0.457 & 0.447 & 0.438 & & 0.914 & 0.490 & 0.486 & 0.477 & & 0.782 & 0.423 & 0.463 & 0.450 \\
MDBH+TPH & 1.496 & 0.977 & 0.939 & $\bm{0.918}$ & & 1.609 & $\bm{0.978}$ & 1.013 & 0.997 & & 1.525 & $\bm{0.920}$ & 0.960 & 0.934 \\
\hline
\hline
$p=0.35$ &  \multicolumn{4}{c}{(I)} & & \multicolumn{4}{c}{(II-a)} & & \multicolumn{4}{c}{(II-b)} \\
 & $M_{1}^{\neq}$ & $M_{2}^{\neq}$  & $M_{3}^{\neq}$ & $M_{4}^{\neq}$ & & $M_{1}^{\neq}$ & $M_{2}^{\neq}$ & $M_{3}^{\neq}$ & $M_{4}^{\neq}$ & & $M_{1}^{\neq}$ & $M_{2}^{\neq}$ & $M_{3}^{\neq}$ & $M_{4}^{\neq}$ \\
\hline
RMSE &  &  &  &  & &  &  &  &  & &  &  &  &  \\
MDBH & 0.884 & 0.707 & 0.680 & 0.682 & & 1.257 & 0.878 & 0.857 & 0.847 & & 1.199 & 0.819 & 0.771 & 0.767 \\
 TPH & 0.943 & 0.673 & 0.604 & 0.607 & & 1.265 & 0.775 & 0.724 & 0.726 & & 1.000 & 0.758 & 0.763 & 0.762 \\
 MDBD+TPH & 1.827 & 1.380 & $\bm{1.284}$ & 1.289 & & 2.522 & 1.653 & 1.581 & $\bm{1.573}$ & & 2.199 & 1.577 & 1.534 & $\bm{1.529}$ \\
 \hline
CRPS &  &  &  &  & &  &  &  &  & &  &  &  &  \\
MDBH & 0.642 & 0.424 & 0.445 & 0.449 & & 0.971 & 0.557 & 0.608 & 0.585 & & 0.992 & 0.523 & 0.504 & 0.497 \\
TPH & 0.707 & 0.390 & 0.406 & 0.400 & & 1.080 & 0.480 & 0.519 & 0.510 & & 1.046 & 0.469 & 0.542 & 0.513 \\
MDBH+TPH & 1.349 & $\bm{0.814}$ & 0.851 & 0.849 & & 2.051 & $\bm{1.037}$ & 1.127 & 1.095 & & 2.038 & $\bm{0.992}$ & 1.046 & 1.010 \\
\hline
\hline
\end{tabular}
}
\end{center}
\label{tab:MT_pred_S1neqS2}
\end{table}

\subsection{$\mathcal{S}_{1} \neq \mathcal{S}_{2}$ and disjoint}

Here, we have two disjoint point patterns, ${\cal S}_1 = \{\bs_1, \bs_2,..., \bs_{n_{1}}\}$ and ${\cal S}_{2} = \{\bs'_{1}, \bs'_{2},...,\bs'_{n_{2}}\}$  with associated responses ${\cal Y}_{1} = \{Y_1(\bs_1), Y_1(\bs_2),...,Y_1(\bs_{n_{1}})\}$ and  ${\cal Y}_{2} = \{Y_2(\bs'_1), Y_2(\bs'_2),...,Y_2(\bs'_{n_{2}})\}$, respectively.  In this setting there is a conceptual $Y_{1}(\bs) \quad \forall \bs \in D$ and a conceptual $Y_{2}(\bs) \quad \forall \bs \in D$.  However, the generative model for $\mathcal{S}_{1}$ is different from that for $\mathcal{S}_{2}$.
Specifically, suppose further that we have environmental vectors $\bX(\bs), \bs \in D$ which we seek to use to explain both the point patterns and the responses. We adopt a LGCP for ${\cal S}_{1}$ with intensity log$\lambda_{1}(\bs) = \bX^{T}(\bs)\balpha_{1} +  \eta_{1}(\bs)$ and a LGCP for ${\cal S}_{2}$ with intensity log$\lambda_{2}(\bs) = \bX^{T}(\bs)\balpha_{2} +  \eta_{2}(\bs)$.  Since the point patterns are imagined as arising under different experiments with sampling settings, there is no reason to make the LGCP's dependent so we assume $\eta_{1}(\bs)$ and $\eta_{2}(\bs)$ are independent GPs.  However, Appendix A demonstrates that, if the responses are modeled as dependent through a bivariate GP for the random effects, then co-kriging for say $Y_{1}(\bs_{0})$ can learn from both ${\cal Y}_{1}$ and  ${\cal Y}_{2}$ even if $Y_1$ and $Y_2$ are never observed jointly.

Only two bivariate process specifications for $(Y_{1}(\bs), Y_{2}(\bs))$ are considered.  One is $M_{3}^{\neq}$, the foregoing coregionalization model, which we denote here as $M_{1}^{*\neq}$ for convenience, i.e., (i) $M_{1}^{*\neq}$: $Y_{1}(\bs) = \bX^{T}(\bs)\bbeta_{1} + a_{11}w_{1}(\bs) + \epsilon_{1}(\bs)$ and  $Y_{2}(\bs) = \bX^{T}(\bs)\bbeta_{2} + a_{21}w_{1}(\bs) + a_{22}w_{2}(\bs) + \epsilon_{2}(\bs)$.  The second is
(ii) $M_{2}^{*\neq}$: $Y_1(\bs) = \bX^{T}(\bs) \bbeta_{1} + \gamma_{1}\eta_{1}(\bs) + a_{11}w_{1}(\bs) + \epsilon_1(\bs)$ and $Y_2(\bs) = \bX^{T}(\bs) \bbeta_{2} + \gamma_{2}\eta_{2}(\bs) + a_{21}w_{1}(\bs) + a_{22}w_{2}(\bs) + \epsilon_2(\bs)$.  That is, we add preferential sampling through shared processes to the coregionalization where the $\gamma$'s are of interest in terms of a PS story.
Given that the LGCP's are independent, we don't consider the possibility that both point pattern intensities can inform about both geostatistical models.
If PS is present, we hope to find significant $\gamma$'s in $M_2^{*\neq}$.  Further, we hope that the inclusion of $\eta_{1}(\bs)$ and $\eta_{2}(\bs)$ in the modeling for $Y_{1}(\bs)$ and $Y_{2}(\bs)$, respectively, will improve predictive performance.
 Perhaps the most important point is that, in the geostatistical setting, $Y_{1}(\bs)$ and $Y_{2}(\bs)$ would not share a GP for their spatial random effects.  However, introducing shared processes from the LGCPs for $\mathcal{S}_{1}$ and $\mathcal{S}_{2}$ enables co-kriging from all of the data.

\subsubsection{A simulation example}

We demonstrate parameter recovery and present model comparison based on the predictive performance for a simulated data example.
The response data  is simulated from model $M_{2}^{=}$, i.e., the shared process model as in Section 3.3.1 but modified so that $\mathcal{S}_{1} \sim \text{LGCP}(\lambda_{1}(\cdot)) $ and $\mathcal{S}_{2}\sim \text{LGCP}(\lambda_{2}(\cdot))$ with $ \log \lambda_{1}(\bs)=\bX^{\top}(\bs)\balpha_{1}+\eta_{1}(\bs)$, $\log \lambda_{2}(\bs)=\bX^{\top}(\bs)\balpha_{2}+\eta_{2}(\bs)$,  $\eta_{1}(\bs) \sim \text{GP}(0, C(||\bs-\bs'||; \btheta_{1})$, and   $\eta_{2}(\bs) \sim \text{GP}(0, C(||\bs-\bs'||; \btheta_{2})$.

The same dimensions are assumed for $D$, again taking the number of grid cells to be $900$.
We also assume a small ($\sigma^2=1/3$) and a large ($\sigma^2=1$) variance for the $\eta_{1}(\bs)$ and $ \eta_{2}(\bs)$ surfaces, again realized at the centroids of the grid cells. The parameter values are:
\begin{align*}
            (\alpha_{1,1}, \alpha_{2,1}) &= (6, 1), \quad (\alpha_{1,2}, \alpha_{2,2}) = (6, 1), \quad \sigma_{1}^2, \sigma_{2}^2 = 1 (\text{or}  1/3), \quad \phi_{1}, \phi_{2} = 3,   \\
            (\beta_{11}, \beta_{21}) &= (0, 0.5), \quad (\beta_{12}, \beta_{12}) = (0, 0.5), \quad (\gamma_{1}, \gamma_{2}) = (1, 0.3) \quad (\tau_{1}^2, \tau_{2}^2) = (0.3, 0.1)
\end{align*}
That is, PS (PS) effects are specified as $\gamma_{1}=1$, a relatively large value and $\gamma_{2}=0.3$, a relatively small value. The total numbers of locations in $\mathcal{S}_{1}$ and $\mathcal{S}_{2}$ are $(n_{1}, n_{2})=(488, 478)$, respectively, for the low variance case, $(n_{1}, n_{2})=(524, 547)$, respectively, for the high variance case.  Weakly informative priors for all parameters are adopted as follows.
\begin{align*}
        \balpha_{1}, \balpha_{2} &\sim \mathcal{N}(\bzero, 100  \bI_{p}), \quad \sigma_{1}^2, \sigma_{2}^2 \sim \mathcal{IG}(2, 0.1), \quad \phi_{1}, \phi_{2}\sim \mathcal{U}(0, 100) \\
        \bbeta_{1}, \bbeta_{2} &\sim \mathcal{N}(\bzero, 100  \bI_{p}), \quad
        \gamma_{1}, \gamma_{2} \sim \mathcal{N}(0, 100), \quad \tau_{1}^2, \tau_{2}^2 \sim \mathcal{IG}(2, 0.1).
\end{align*}

Table \ref{tab:sim_pred_S1disS2} shows the predictive performance results. For (I), $M_1^{*\neq}$ and $M_2^{*\neq}$ reveal similar performance with RMSE and CRPS. Again, this is a reasonable result because the true surface by the shared process model is recovered with the coregionalization model under the random holdout case. For (II), $M_2^{*\neq}$ shows better performance with CRPS, though the scores with RMSE of $M_1^{*\neq}$ and $M_2^{*\neq}$ are similar.
Table \ref{tab:sim_est_S1disS2} shows the estimation results for both models.  The relevant parameters for the simulation model are recovered well by $M_2^{*\neq}$, especially the $\gamma$'s are well estimated by $M_2^{*\neq}$ and $a_{2,1}$ is insignificant.

\begin{table}[htbp]
\caption{Model comparison of predictive performance for the simulated data in the disjoint $\mathcal{S}_{1}\neq \mathcal{S}_{2}$ case for low ($\sigma^2=1/3$) and high ($\sigma^2=1$) variance. The bold fonts suggest the best performance for each holdout}
\begin{center}
\scalebox{0.80}{
\begin{tabular}{lcccccccccccccc}
\hline
\hline
low &  \multicolumn{2}{c}{(I)} & & \multicolumn{2}{c}{(II-a)} & & \multicolumn{2}{c}{(II-b)} \\
   & $M_{1}^{*\neq}$ & $M_{2}^{*\neq}$ & & $M_{1}^{*\neq}$ & $M_{2}^{*\neq}$ & & $M_{1}^{*\neq}$ & $M_{2}^{*\neq}$ \\
\hline
 $\text{RMSE}_{1}$  & 0.611 & 0.615 & & 0.594 & 0.575 & & 0.609 & 0.608 \\
 $\text{RMSE}_{2}$  & 0.316 & 0.317 & & 0.314 & 0.311 & & 0.349 & 0.349 \\
  $\text{RMSE}_{1+2}$ & $\bm{0.927}$ & 0.932 & & 0.908 & $\bm{0.886}$ & & 0.958 & $\bm{0.957}$ \\
 \hline
$\text{CRPS}_{1}$ & 0.405 & 0.401 &  & 0.403 & 0.376 &  & 0.403 & 0.388 \\
 $\text{CRPS}_{2}$ & 0.204 & 0.209 & & 0.205 & 0.203 & & 0.243 & 0.250 \\
 $\text{CRPS}_{1+2}$ & $\bm{0.609}$ & 0.610 & & 0.608 & $\bm{0.579}$ & & 0.646 & $\bm{0.638}$ \\
\hline
\hline
high &  \multicolumn{2}{c}{(I)} & & \multicolumn{2}{c}{(II-a)} & & \multicolumn{2}{c}{(II-b)} \\
 & $M_{1}^{*\neq}$ & $M_{2}^{*\neq}$ & & $M_{1}^{*\neq}$ & $M_{2}^{*\neq}$ & & $M_{1}^{*\neq}$ & $M_{2}^{*\neq}$ \\
\hline
 $\text{RMSE}_{1}$ & 0.692 & 0.721 & & 0.698 & 0.667 & & 0.696 & 0.704 \\
 $\text{RMSE}_{2}$ & 0.320 & 0.321 & & 0.321 & 0.322 & & 0.320 & 0.316 \\
 $\text{RMSE}_{1+2}$ & $\bm{1.012}$ & 1.042 & & 1.019 & $\bm{0.989}$ & & $\bm{1.016}$ & 1.020 \\
 \hline
$\text{CRPS}_{1}$ & 0.455 & 0.455 & & 0.460 & 0.412 & & 0.456 & 0.440 \\
 $\text{CRPS}_{2}$ & 0.206 & 0.207 & & 0.207 & 0.206 & & 0.227 & 0.217 \\
 $\text{CRPS}_{1+2}$  & $\bm{0.661}$ & 0.662 & & 0.667 & $\bm{0.618}$ & & 0.683 & $\bm{0.657}$ \\
\hline
\hline
\end{tabular}
}
\end{center}
\label{tab:sim_pred_S1disS2}
\end{table}

\section{Preferential sampling and dependence bias}

Here, we demonstrate that, under bivariate PS, we can be misled with regard to the dependence between the two responses at a location and, more generally, the behavior of the cross-covariance function.  That is, with sampling bias in bivariate response data, we can investigate the impact on second moment behavior for the bivariate process driving the data.  Such sampling bias can arise in practice if sampling locations favor pairs which tend to display positive association or pairs which tend to display negative association.  For this investigation we work with the $\mathcal{S}_{1} = \mathcal{S}_{2}$ case; also, there is no need to bring in missing data for this discussion.  However, following the  calculations in Section 3.1, we have the cross-covariance specification for models $M_{3}^{=}$ and $M_{4}^{=}$. The cross covariance matrix for $M_{3}^{=}$, $C_{M_{3}^{=}}(\bs -\bs')=$\\ $\rho_{\omega_{1}}(\bs-\bs')\left(
                                                                                    \begin{array}{cc}
                                                                                      a_{11}^{2} & a_{11}a_{21} \\
                                                                                      a_{11}a_{21} & a_{21}^{2} \\
                                                                                    \end{array}
                                                                                  \right)$ + $\rho_{\omega_{2}}(\bs-\bs')\left(
                                                                                    \begin{array}{cc}
                                                                                      0 & 0 \\
                                                                                      0 & a_{22}^{2} \\
                                                                                    \end{array}
                                                                                  \right)$.\\

Under $M_{4}^{=}$, $C_{M_{4}^{=}}(\bs -\bs')=$   \\
 $c_{\eta}(\bs - \bs')\left(
                             \begin{array}{cc}
                               \gamma_{1}^{2} & \gamma_{1}\gamma_{2} \\
                               \gamma_{1}\gamma_{2} & \gamma_{2}^{2} \\
                             \end{array}
                           \right)$ + $\rho_{\omega_{1}}(\bs-\bs')\left(
                                                                                    \begin{array}{cc}
                                                                                      a_{11}^{2} & a_{11}a_{21} \\
                                                                                      a_{11}a_{21} & a_{21}^{2} \\
                                                                                    \end{array}
                                                                                  \right)$ + $\rho_{\omega_{2}}(\bs-\bs')\left(
                                                                                    \begin{array}{cc}
                                                                                      0 & 0 \\
                                                                                      0 & a_{22}^{2} \\
                                                                                    \end{array}
                                                                                  \right)$.  \\

The goal is to demonstrate the difference in inference regarding dependence structure between these two models and, specifically, to show the bias that can arise under $M_{3}^{=}$ and how $M_{4}^{=}$ changes the inference.  We can examine the results at any particular location and also make a comparison of the cross-covariance between locations.  We simulate under $M_{4}^{=}$ to enable positive or negative correlations.  That is, we introduce correlation bias through the sampling of the $(Y_{1}(\bs), Y_{2}(\bs))$ pairs, changing the strength of association, perhaps changing the sign of the correlation.  It is clear how to draw a sample of $(Y_{1}(\bs), Y_{2}(\bs))$ pairs from $\mathcal{S}$ to encourage positive or negative correlation. Then, we can obtain the raw correlation from the biased data to compare with the revised posterior correlation under $M_{4}^{=}$ as well as the true correlation under $M_{4}^{=}$.

Here, we are not concerned with the prediction bias; we just seek to reveal the dependence bias. There is no model checking or comparison.  In Section 4.1 we illustrate with a simulation example.

\subsection{A simulation example of dependence bias under preferential sampling}

We examine inference on the dependence structure through shared processes and correlated Gaussian processes. We generate data using the model $M_{4}^{=}$ in Section 3.1, but, in order to simplify the discussion, we exclude covariates.
We will examine whether the dependence structure can be recovered for two datasets, data1 with a positive $\gamma_{2}$ and data2 with a negative value in order to capture positive and negative strengths of correlation.  Specifically, we have
\begin{align}
        \mathcal{S}&\sim \text{LGCP}(\lambda(\cdot)) \nonumber \\
        \log \lambda(\bs)&=\alpha_{1}+\eta(\bs), \quad \eta(\cdot) \sim \text{GP}(0, C(||\bs-\bs'||; \btheta) \nonumber \\        Y_{1}(\bs)&=\gamma_{1}\eta(\bs)+a_{11}w_{1}(\bs)+\epsilon_{1}(\bs), \quad  \epsilon_{1}(\bs)\sim \mathcal{N}(0, \tau_{1}^2)  \nonumber \\
        Y_{2}(\bs)&=\gamma_{2}\eta(\bs)+a_{21}w_{1}(\bs)+a_{22}w_{2}(\bs)+\epsilon_{2}(\bs), \quad \epsilon_{2}(\bs)\sim \mathcal{N}(0, \tau_{2}^2). \nonumber \\
        w_{1}(\cdot) &\sim \text{GP}(0, C(||\bs-\bs'||; \btheta_{w1}), \quad w_{2}(\cdot) \sim \text{GP}(0, C(||\bs-\bs'||; \btheta_{w2})
\end{align}
The parameter settings for data1 are:
           $ \alpha_{1} = 6, \quad \sigma^2, a_{11}^2, a_{22}^2 = 1, \quad \phi, \phi_{w1}, \phi_{w2} = 1, \quad (\gamma_{1}, \gamma_{2}) = (1, 0.3) \quad (\tau_{1}^2, \tau_{2}^2) = (0.3, 0.1), a_{21} = -0.4$.
For data2, we change the sign of $\gamma_{2}$, i.e., $\gamma_{2}=-0.3$.
The simulated number of locations is 624. In addition, we introduce artificial sampling biases by sampling 70$\%$ of all locations according to the descending order of the $Y_{1}$, this sampling is expected to distort the covariance structure between $Y_{1}$ and $Y_{2}$. Although the data are simulated by $M_{4}^{=}$, the estimation by $M_{4}^{=}$ doesn't necessarily cover the true covariance structure with the bias sampling.
To implement model $M_{3}^{=}$ and $M_{4}^{=}$, we assume the same prior settings and MCMC fitting as above.

\begin{figure}[ht]
  \begin{center}
   \includegraphics[width=16cm]{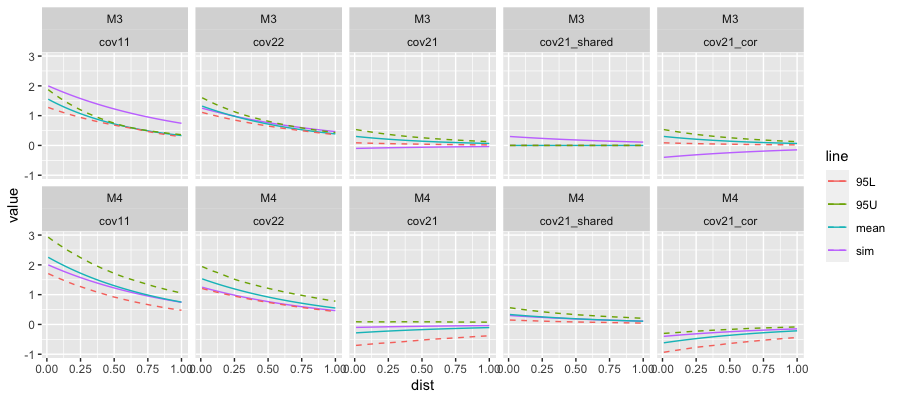}
  \end{center}
  \caption{The plots of the true covariance surfaces and their posterior means and $95\%$ credible intervals for data1: cov$(Y_{1}(\bs), Y_{1}(\bs'))$ (\texttt{cov11}), cov$(Y_{2}(\bs), Y_{2}(\bs'))$ (\texttt{cov22}), cov$(Y_{1}(\bs), Y_{2}(\bs'))$ (\texttt{cov21}), cov$(\gamma_{1} \eta(\bs), \gamma_{2} \eta(\bs'))$ (\texttt{cov21$\_$shared}) and cov$(a_{1, 1}w_{1}(\bs), a_{2, 1}w_{1}(\bs'))$ (\texttt{cov21$\_$corr}). }
  \label{fig:sim5_covariance_dependence}
\end{figure}

\begin{figure}[ht]
  \begin{center}
   \includegraphics[width=16cm]{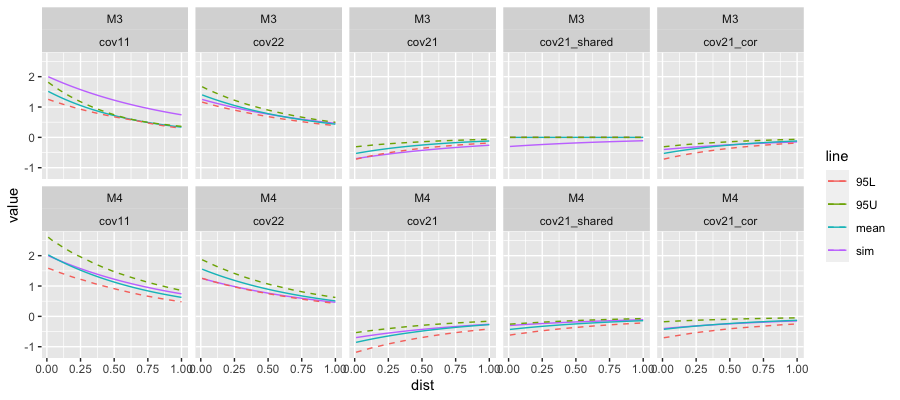}
  \end{center}
  \caption{The plots of the true covariance surfaces and their posterior means and $95\%$ credible intervals for data2: cov$(Y_{1}(\bs), Y_{1}(\bs'))$ (\texttt{cov11}), cov$(Y_{2}(\bs), Y_{2}(\bs'))$ (\texttt{cov22}), cov$(Y_{1}(\bs), Y_{2}(\bs'))$ (\texttt{cov21}), cov$(\gamma_{1} \eta(\bs), \gamma_{2} \eta(\bs'))$ (\texttt{cov21$\_$shared}) and cov$(a_{1, 1}w_{1}(\bs), a_{2, 1}w_{1}(\bs'))$ (\texttt{cov21$\_$corr}). }
  \label{fig:sim7_covariance_dependence}
\end{figure}

Figures \ref{fig:sim5_covariance_dependence} and \ref{fig:sim7_covariance_dependence} show the simulated covariance surfaces and their posterior means and 95$\%$ credible intervals at all locations for data1 and data2.
The 95th percentile credible intervals will be affected by the sample size, but here, for illustration, the sample size is adjusted to essentially agree with that of the  MDBH/TPH data.  Altogether, covariance structure is estimated well by $M_{4}^{=}$. Especially, for cov$(Y_{1}(\bs), Y_{2}(\bs'))$, the covariance contributions from the shared process and the coregionalized GPs are distinguished.
However, $M_{3}^{=}$ estimates cov$(Y_{1}(\bm{s}), Y_{1}(\bm{s}'))$ and cov$(Y_{1}(\bm{s}), Y_{2}(\bm{s}'))$ with biases. Figures \ref{fig:sim5_covariance_dependence_biased} and \ref{fig:sim7_covariance_dependence_biased} show the simulated covariance surfaces and their posterior means and 95$\%$ credible intervals at locations with biases for data1 and data2. $M_{4}^{=}$ doesn't capture cov$(Y_{1}(\bm{s}), Y_{1}(\bm{s}'))$ and cov$(Y_{1}(\bm{s}), Y_{2}(\bm{s}'))$ surfaces, but
$M_{4}^{=}$ shows wider intervals than $M_{3}^{=}$ and also captures cov$(Y_{1}(\bm{s}), Y_{1}(\bm{s}'))$ and cov$(Y_{1}(\bm{s}), Y_{2}(\bm{s}'))$ without bias.
Figures \ref{fig:sim5_marginalcov_dependence} and \ref{fig:sim7_marginalcov_dependence} present the posterior means and 95$\%$ credible intervals for the covariance and correlation at the same locations for data1 and data2. For both figures, $M_{4}^{=}$ shows wider intervals for cov$(Y_{1}(\bs), Y_{1}(\bs))$ and captures the true value even with the bias sampling while $M_{3}^{=}$ fails to capture it.
For cov$(Y_{1}(\bs), Y_{2}(\bs))$, both models capture the true value for data1.  However, $M_{3}^{=}$ fails to capture it for data2 while $M_{4}^{=}$ estimates it but with slightly wider intervals.

\section{Summary}

We have proposed novel extensions of the notion of PS to the context of a pair of response variables.  As with customary PS, the goals are to recognize the presence of PS effects and to examine improved spatial prediction if there are.  Using both simulation and real data we have demonstrated, through shared process modeling, the ability to identify PS effects and also to improve co-kriging in the presence of these effects.  Further, we have shown that PS can bias spatial dependence structure but, again, a shared process specification can clarify the true dependence behavior.

In the literature, there are many models for dependent spatial point patterns, e.g., clustering and inhibition specifications \citep{GelfandSchliep(18)}.  However, in order to supply a shared process specification under geostatistical modeling, employing GPs is most convenient. In this regard, recent work \citep{Virhsetal(21)} introduces spatial aggregation to a Gibbs process using a GP and offers the possibility of further PS investigation.  Another path for future work involves spatio-temporal response data collection, opening the potential of spatial bias varying over time in the data collection. This would lead to space-time geostatistical modeling and space-time point pattern intensities using shared space-time GPs.

\bibliographystyle{chicago}
\bibliography{SP}

\newpage
\appendix
\section{Co-kriging with response variables at disjoint locations}

Here, we clarify that, with dependent spatial responses, using geostatistical modeling, co-kriging for say $Y_{1}(\bs_0)$ within a Bayesian framework will depend on both $\mathcal{Y}_{1}$ and $\mathcal{Y}_{2}$ even if the set of locations $\mathcal{S}_{1}$ where $\mathcal{Y}_{1}$ was observed and the set of locations where $\mathcal{Y}_{2}$ was observed are disjoint.
Suppose the bivariate geostatistical setting
\begin{equation*}
[\mathcal{Y}_{1}|\mathcal{S}_1, \bet_1, \bbeta_1, \tau_{1}^{2}][\mathcal{Y}_{2}|\mathcal{S}_2, \bet_2, \bbeta_{2}, \tau_{2}^{2}][\bet_1, \bet_2|\btheta][\bbeta_1, \bbeta_2, \tau_{1}^{2}, \tau_{2}^{2}, \btheta].
\end{equation*}
Here, $Y_{1}(\bs)$ is a standard geostatistical model with regression coefficients $\bbeta_1$, GP $\eta_1(\bs)$, and nugget $\tau_{1}^{2}$ and $Y_{2}(\bs)$ is a standard geostatistical model with regression coefficients $\bbeta_2$, GP $\eta_2(\bs)$, and nugget $\tau_{2}^{2}$.  $(\eta_{1}(\bs), \eta_{2}(\bs))$ follow a bivariate spatial process (perhaps a bivariate GP) with parameters $\btheta$.  So, $Y_1(\bs)$ and $Y_{2}(\bs)$ are a bivariate process model with a general cross-covariance dependence structure incorporated into the joint distribution $[\bet_{1}, \bet_{2}|\btheta]$.

Consider the posterior predictive distribution, $[Y_{1}(\bs_0)|\mathcal{Y}_{1}, \mathcal{Y}_{2}]$.  We show that it does depend on $\mathcal{Y}_{2}$. $[Y_{1}(\bs_0)|\mathcal{Y}_{1}, \mathcal{Y}_{2}]= \int [Y_{1}(\bs_0)|\eta_{1}(\bs_0), \bbeta_{1}, \tau_{1}^{2}][\eta_{1}(\bs_0), \bbeta_{1}, \tau_{1}^{2}| \mathcal{Y}_{1}, \mathcal{Y}_{2}]$.

However, $[\eta_{1}(\bs_0), \bbeta_{1}, \tau_{1}^{2}| \mathcal{Y}_{1}, \mathcal{Y}_{2}] \propto \int [\eta_{1}(\bs_0)|\bet_1, \bet_2, \btheta][\bet_1, \bet_2, \btheta, \bbeta_1, \tau_{1}^{2}|\mathcal{Y}_{1}, \mathcal{Y}_{2}]d\bet_{1}d\bet_{2}d\btheta$.
The integral becomes $\int g_{1}(\mathcal{Y}_{1}, \bet_{1}, \bbeta_{1}, \tau_{1}^{2})g_{2}(\mathcal{Y}_{2}, \bet_{2}) [\bet_1, \bet_2, \btheta][\btheta]d\bet_{1}d\bet_{2}d\btheta$, after a little manipulation.  This integration yields $g( \eta_{1}(\bs_0), \bbeta_{1} \tau_{1}^{2}, \mathcal{Y}_{1}, \mathcal{Y}_{2})$ due to the dependence between $\bet_{1}$ and $\bet_{2}$.  Marginalizing over $\bbeta_{1}$ and $\tau_{1}^{2}$ shows that the posterior predictive distribution for $Y_{1}(\bs_0)$ given $\mathcal{Y}_{1}, \mathcal{Y}_{2}$ does depend on $\mathcal{Y}_{2}$.  Hence co-kriging under model $M_3^{\neq}$ utilizes all of the data even when the point patterns are disjoint.  Evidently, this carries over to $M_4^{\neq}$ when a shared process is added.

\newpage
\counterwithin{table}{section}
\counterwithin{figure}{section}
\section{Figures}
\begin{figure}[ht]
  \begin{center}
   \includegraphics[width=15cm]{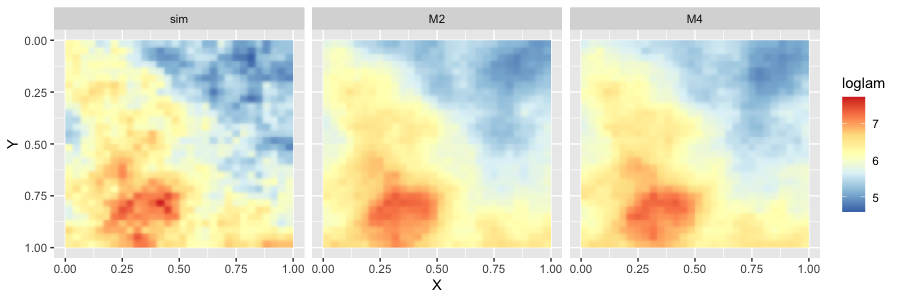}
  \end{center}
    \begin{center}
   \includegraphics[width=15cm]{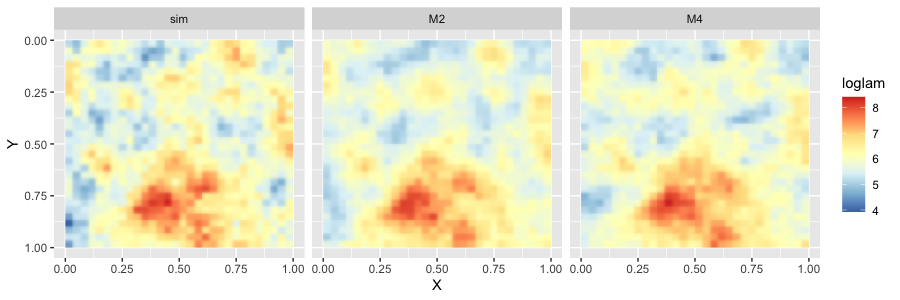}
  \end{center}
\caption{log $\lambda$ surface for the low variance case (upper) and the high variance case (bottom): the simulated true surface (left), the posterior mean surface of $M_{2}^{=}$ (middle) and the posterior mean surface of $M_{4}^{=}$ (right).}
  \label{fig:sim_int_S1=S2}
\end{figure}

\begin{figure}[ht]
  \begin{center}
   \includegraphics[width=16cm]{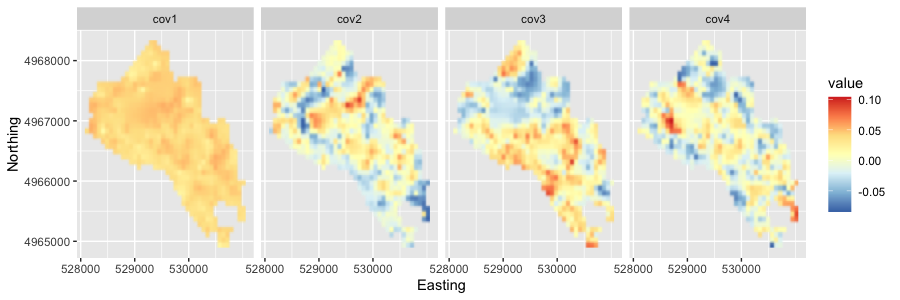}
  \end{center}
  \caption{The plots of the four covariate surfaces on all locations.}
  \label{fig:MT_cov}
\end{figure}

\begin{figure}[ht]
  \begin{center}
   \includegraphics[width=15cm]{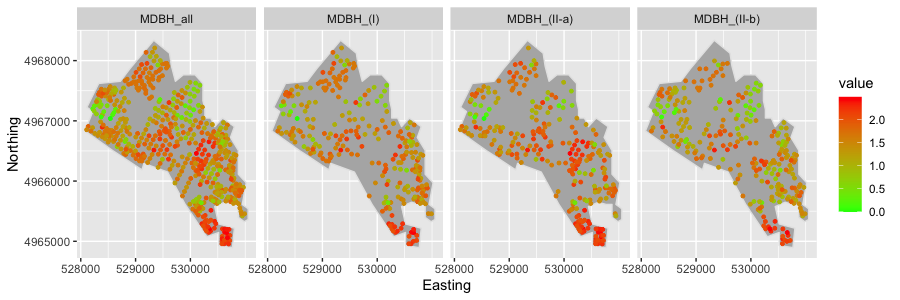}
  \end{center}
  \begin{center}
   \includegraphics[width=15cm]{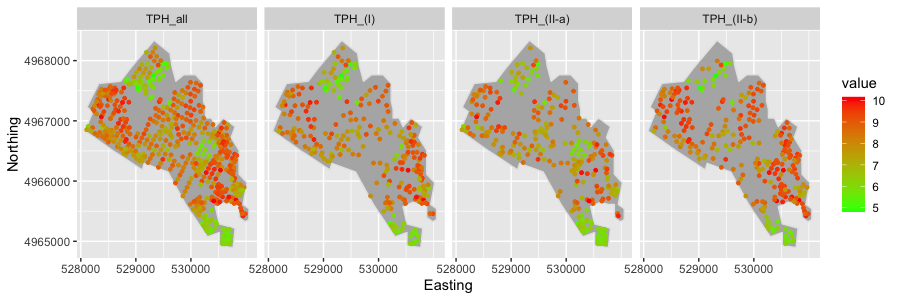}
  \end{center}
  \caption{The plots of MDBH (top) and TPH (bottom) on all locations (left) and locations preserved by each holdout strategy for $\mathcal{S}_{1}\neq \mathcal{S}_{2}$. (I):random, (II-a):partially overlapped with the descending order of MDBH, (II-b):partially overlapped with the descending order of TPH}
  \label{fig:MT_Y_S1neqS2}
\end{figure}

\begin{figure}[ht]
  \begin{center}
   \includegraphics[width=16cm]{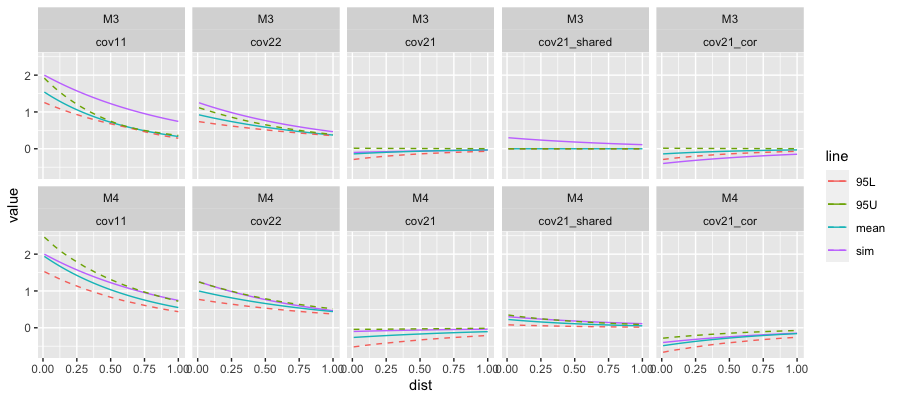}
  \end{center}
  \caption{The plots of the true covariance surfaces and their posterior means and $95\%$ credible intervals for the bias sampling case for data1: cov$(Y_{1}(\bs), Y_{1}(\bs'))$ (\texttt{cov11}), cov$(Y_{2}(\bs), Y_{2}(\bs'))$ (\texttt{cov22}), cov$(Y_{1}(\bs), Y_{2}(\bs'))$ (\texttt{cov21}), cov$(\gamma_{1} \eta(\bs), \gamma_{2} \eta(\bs'))$ (\texttt{cov21$\_$shared}) and cov$(a_{1, 1}w_{1}(\bs), a_{2, 1}w_{1}(\bs'))$ (\texttt{cov21$\_$corr}). }
  \label{fig:sim5_covariance_dependence_biased}
\end{figure}

\begin{figure}[ht]
  \begin{center}
   \includegraphics[width=16cm]{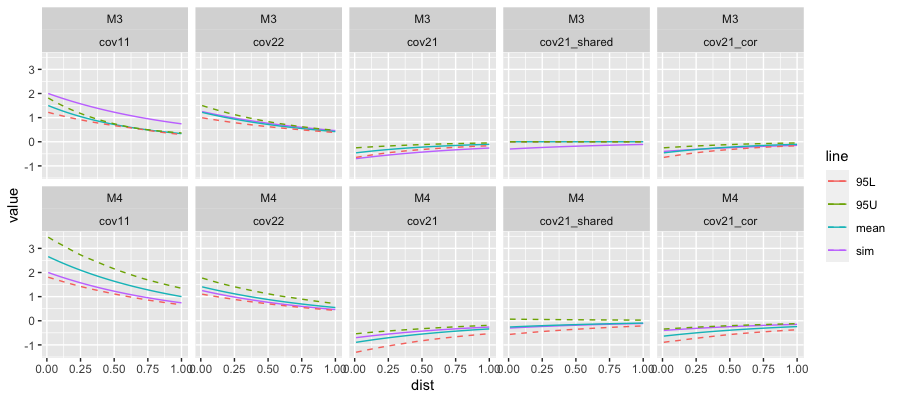}
  \end{center}
  \caption{The plots of the true covariance surfaces and their posterior means and $95\%$ credible intervals for the bias sampling case for data2: cov$(Y_{1}(\bs), Y_{1}(\bs'))$ (\texttt{cov11}), cov$(Y_{2}(\bs), Y_{2}(\bs'))$ (\texttt{cov22}), cov$(Y_{1}(\bs), Y_{2}(\bs'))$ (\texttt{cov21}), cov$(\gamma_{1} \eta(\bs), \gamma_{2} \eta(\bs'))$ (\texttt{cov21$\_$shared}) and cov$(a_{1, 1}w_{1}(\bs), a_{2, 1}w_{1}(\bs'))$ (\texttt{cov21$\_$corr}). }
  \label{fig:sim7_covariance_dependence_biased}
\end{figure}

\begin{figure}[ht]
  \begin{center}
   \includegraphics[width=16cm]{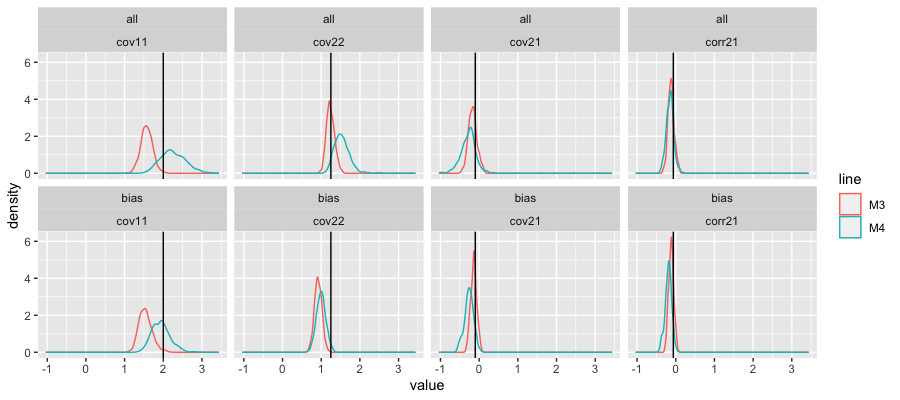}
  \end{center}
  \caption{The plots of the posterior means and $95\%$ credible intervals for local covariance and correlation for all samples (top) and the bias sampling (bottom) for data1, the black vertical line suggests the true value: cov$(Y_{1}(\bs), Y_{1}(\bs))$ (\texttt{cov11}), cov$(Y_{2}(\bs), Y_{2}(\bs))$ (\texttt{cov22}), cov$(Y_{1}(\bs), Y_{2}(\bs))$ (\texttt{cov21}) and corr$(Y_{1}(\bs), Y_{2}(\bs))$ (\texttt{corr21}). }
  \label{fig:sim5_marginalcov_dependence}
\end{figure}

\begin{figure}[ht]
  \begin{center}
   \includegraphics[width=16cm]{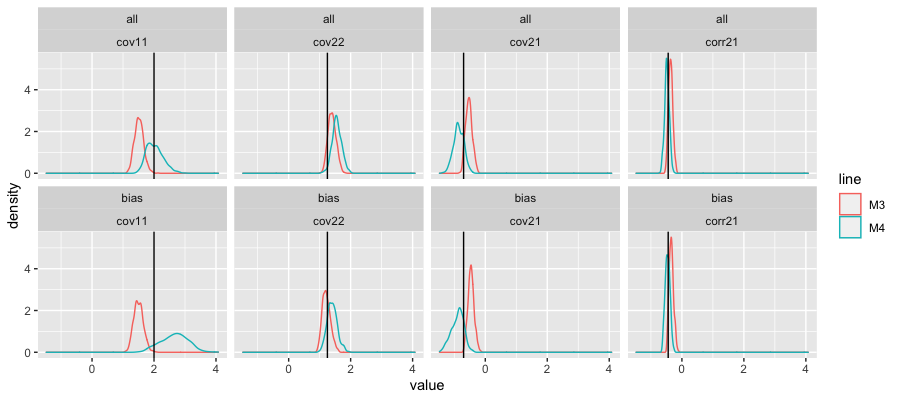}
  \end{center}
  \caption{The plots of the posterior means and $95\%$ credible intervals for local covariance and correlation for all samples (top) and the bias sampling (bottom) for data2, the black vertical line suggests the true value: cov$(Y_{1}(\bs), Y_{1}(\bs))$ (\texttt{cov11}), cov$(Y_{2}(\bs), Y_{2}(\bs))$ (\texttt{cov22}), cov$(Y_{1}(\bs), Y_{2}(\bs))$ (\texttt{cov21}) and corr$(Y_{1}(\bs), Y_{2}(\bs))$ (\texttt{corr21}). }
  \label{fig:sim7_marginalcov_dependence}
\end{figure}

\newpage
\section{The details of estimation results}

\begin{table}[htbp]
\caption{The estimation results for the simulated data in the $\mathcal{S}_{1}=\mathcal{S}_{2}$ case for low ($\sigma^2=1/3$) and high ($\sigma^2=1$) variance.}
\begin{center}
\scalebox{0.75}{
\begin{tabular}{lccccccccccc}
\hline
\hline
low &  & \multicolumn{2}{c}{$M_{1}^{=}$} & \multicolumn{2}{c}{$M_{2}^{=}$} & \multicolumn{2}{c}{$M_{3}^{=}$} & \multicolumn{2}{c}{$M_{4}^{=}$} \\
   & true & Mean &  $95\%$ Int  &  Mean &  $95\%$ Int & Mean &  $95\%$ Int  & Mean &  $95\%$ Int \\
\hline
 $\alpha_{1}$ & 6 & -  & - & 5.905 & [5.547, 6.257] & - & - & 5.692 & [4.614, 6.807] \\
 $\alpha_{2}$ & 1 & -  & - & 1.125 & [0.029, 2.025] & - & - & 0.878 & [-0.574, 2.190] \\
 $\beta_{1, 1}$ & 0 & 0.210  & [0.147, 0.271] & -0.089 & [-0.426, 0.241] & -0.413 & [-1.668, 0.952] & 0.044 & [-2.098, 1.966] \\
 $\beta_{1, 2}$ & 0.5 & 0.719 & [0.509, 0.955] & 0.561 & [-0.482, 1.377] & 0.476 & [-0.967, 1.814] & 0.343 & [-1.083, 1.479] \\
 $\beta_{2, 1}$ & 0 & 0.034 & [0.003, 0.069] & -0.064 & [-0.183, 0.048] & -0.159 & [-0.536, 0.277] & -0.139 & [-0.527, 0.226] \\
 $\beta_{2, 2}$ & 0.5 & 0.646 & [0.535, 0.763] & 0.588 & [0.241, 0.875] & 0.565 & [0.085, 1.012] & 0.506 & [-0.001, 0.932] \\
 $\gamma_{1}$ & 1 & - & - & 0.954 & [0.746, 1.238] & - & - & 0.949 & [0.713, 1.241] \\
 $\gamma_{2}$ & 0.3 & - & - & 0.317 & [0.228, 0.426] & - & - & 0.331 & [0.244, 0.442] \\
 $a_{1, 1}$ & - &- & - & - & - & 0.078 & [0.021, 0.191] & 5.472 & [0.161, 16.969] \\
 $a_{2, 1}$ & - &- & - & - & - & 1.328 & [0.415, 3.791] & -0.012 & [-0.748, 0.676] \\
 $a_{2, 2}$ & - &- & - & - & - &  0.429 & [0.141, 1.196] & 0.006 & [0.000, 0.018] \\
 $a_{1,1}^2\phi_{w1}$ & - & -  & - & - & - & 0.000 & [0.000, 0.003] & 0.005 & [0.000, 0.035] \\
 $a_{2,2}^2\phi_{w2}$ & - & -  & - & - & - & 0.078 & [0.033, 0.173] & 0.000 & [0.000, 0.000] \\
 $\sigma_{\eta}^2\phi_{\eta}$ & 1 &-  & - & 0.804 & [0.387, 1.403] & - & - & 0.890 & [0.455, 1.504] \\
 $\tau_{1}^2$ & 0.3 & 0.475  & [0.420, 0.538] & 0.302 & [0.264, 0.351] & 0.298 & [0.256, 0.349] & 0.299 & [0.260, 0.345]\\
 $\tau_{2}^2$ & 0.1 & 0.119  & [0.105, 0.135] & 0.100 & [0.087, 0.114] & 0.100 & [0.086, 0.115] & 0.098 & [0.086, 0.110]\\
\hline
\hline
high &  &\multicolumn{2}{c}{$M_{1}^{=}$} &  \multicolumn{2}{c}{$M_{2}^{=}$} & \multicolumn{2}{c}{$M_{3}^{=}$} & \multicolumn{2}{c}{$M_{4}^{=}$} \\
   & true & Mean &  $95\%$ Int  & Mean &  $95\%$ Int & Mean &  $95\%$ Int & Mean &  $95\%$ Int  \\
\hline
 $\alpha_{1}$ & 6 & -  & - & 5.010 & [5.712, 6.579] & - & - & 5.998 & [5.629, 6.393]\\
 $\alpha_{2}$ & 1 & -  & - & 0.378 & [-0.667, 1.526] & - & - & 0.659 & [-0.401, 1.637] \\
 $\beta_{1, 1}$ & 0 & 0.468 & [0.386, 0.546] & 0.010 & [-0.296, 0.597] & 0.274 & [-0.731, 1.428] & -0.855 &[-1.219, 0.874] \\
 $\beta_{1, 2}$ & 0.5 & 0.714 & [0.442, 0.980] & -0.474 & [-1.559, 0.674] & -0.338 & [-1.768, 0.998] & -0.322 & [-1.353, 0.659]\\
 $\beta_{2, 1}$ & 0 & 0.123 & [0.092, 0.154] & -0.001 & [-0.091, 0.156] & 0.053 & [-0.105, 0.207] & -0.012 & [-0.210, 0.144] \\
 $\beta_{2, 2}$ & 0.5 & 0.553 & [0.446, 0.652] & 0.237 & [-0.055, 0.559] & 0.265 & [-0.135, 0.634] & 0.270 & [-0.035, 0.548]\\
 $\gamma_{1}$ & 1 & -  & - & 1.037 & [0.884, 1.232] & - & - & 1.082 & [0.911, 1.279] \\
 $\gamma_{2}$ & 0.3 & - & - & 0.281 & [0.230, 0.336] & - & - & 0.294 & [0.234, 0.357]\\
 $a_{1, 1}$ & - & -  & - & - & - & 0.553 & [0.116, 1.481] & 0.051 & [0.000, 0.181] \\
 $a_{2, 1}$ & - & -  & - & - & - & 0.884 & [0.594, 1.630] & 0.000 & [-0.144, 0.154] \\
 $a_{2, 2}$ & - & -  & -  & - & - & 0.244 & [0.159, 0.447] & 0.138 & [0.009, 0.419] \\
 $a_{1,1}^2\phi_{w1}$ & - & -  & - & - & - & 0.015 & [0.001, 0.096] & 0.006 & [0.000, 0.063] \\
 $a_{2,2}^2\phi_{w2}$ & - & -  & - & - & - & 0.223 & [0.138, 0.329] & 0.000 & [0.000, 0.005] \\
 $\sigma_{\eta}^2\phi_{\eta}$ & 3 & -  & - & 3.255 & [2.183, 4.619] & - & - & 2.595 & [1.558, 3.825] \\
 $\tau_{1}^2$ & 0.3 & 0.831 & [0.744, 0.935] & 0.335 & [0.286, 0.388] & 0.346 & [0.292, 0.403] & 0.339 & [0.279, 0.400] \\
 $\tau_{2}^2$ & 0.1 & 0.126 & [0.112, 0.141] & 0.089 & [0.077, 0.100] & 0.088 & [0.077, 0.100] & 0.089 & [0.078, 0.102] \\
\hline
\hline
\end{tabular}
}
\end{center}
\label{tab:sim_est_S1=S2}
\end{table}

\begin{table}[ht]
\caption{The estimation results for the MDBH-TPH data on all locations in the $\mathcal{S}_{1}=\mathcal{S}_{2}$ case.}
\begin{center}
\scalebox{0.75}{
\begin{tabular}{lcccccccccc}
\hline
\hline
 & \multicolumn{2}{c}{$M_{1}^{=}$} & \multicolumn{2}{c}{$M_{2}^{=}$} & \multicolumn{2}{c}{$M_{3}^{=}$} & \multicolumn{2}{c}{$M_{4}^{=}$} \\
    & Mean &  $95\%$ Int  &  Mean &  $95\%$ Int & Mean &  $95\%$ Int  & Mean &  $95\%$ Int \\
\hline
 $\alpha_{0}$  & -  & - & 6.884 & [6.447, 7.331] & - & - & 6.378 & [5.998, 6.768] \\
 $\alpha_{1}$  & -  & - & -6.045 & [-11.24, -1.626] & - & - & -1.196 & [-8.485, 6.895] \\
 $\alpha_{2}$  & -  & - & -0.214 & [-2.977, 2.325] & - & - & -0.570 & [-4.560, 2.782] \\
 $\alpha_{3}$  & -  & - & 0.165 & [-1.932, 2.558] & - & - & 0.399 & [-1.789, 2.425] \\
 $\alpha_{4}$  & -  & - & 1.321 & [-1.399, 4.225] & - & - & 0.252 & [-1.789, 2.425] \\
 $\beta_{1, 0}$  & -0.470 & [-1.134, 0.183] & 1.572 & [-0.283, 4.078] & -0.758 & [-1.442, -0.024] & -0.482 & [-1.102, 0.048] \\
 $\beta_{1, 1}$  & 9.864 & [-5.406, 25.46] & 24.46 & [13.06, 35.63] & 17.52 & [2.735, 31.61] & 12.90 & [2.579, 22.78] \\
 $\beta_{1, 2}$  & 1.487 & [-1.551, 4.548] & -0.014 & [-1.638, 1.657] & 0.928 & [-1.350, 3.272] & 0.687 & [-0.941, 2.280] \\
 $\beta_{1, 3}$  & 11.76 & [7.950, 15.42] & 11.44 & [9.105, 13.85] & 9.810 & [6.506, 12.93] & 8.124 & [5.734, 10.54] \\
 $\beta_{1, 4}$  & -1.250 & [-4.081, 1.415] & -2.160 & [-3.834, -0.537] & -1.402 & [-3.567, 0.765] & -2.455 & [-4.017, -0.974] \\
 $\beta_{2, 0}$  & 0.364 & [-0.335, 1.078] & -2.854 & [-5.912, -0.676] & 0.164 & [-0.479, 0.761] & 0.252 & [-0.257, 0.763] \\
 $\beta_{2, 1}$  & -7.364 & [-24.10, 8.940] & -9.091 & [-19.51, 2.004] & -5.150 & [-18.15, 8.091] & -9.974 & [-19.75, -0.798] \\
 $\beta_{2, 2}$  & -2.406 & [-5.677, 0.899] & -1.643 & [-3.049, -0.168] & -2.180 & [-4.348, -0.217] & -1.785 & [-3.265, -0.237] \\
 $\beta_{2, 3}$  & -5.476 & [-9.436, -1.570] & -1.779 & [-3.959, 0.380] & -2.415 & [-5.107, 0.449] & -3.237 & [-5.501, -1.003] \\
 $\beta_{2, 4}$  & 1.274 & [-1.651, 4.008] & 2.277 & [0.817, 3.763] & 0.412 & [-1.707, 2.560] & 1.748 & [0.330, 3.152] \\
 $\gamma_{1}$  & - & - & 7.652 & [4.669, 11.17] & - & - & 1.163 & [0.379, 2.093] \\
 $\gamma_{2}$  & - & - & -9.436 & [-13.92, -5.808] & - & - & -1.116 & [-2.019, -0.250] \\
 $a_{1, 1}$  &- & - & - & - & 0.637 & [0.425, 1.207] & 0.706 & [0.480, 1.340] \\
 $a_{2, 1}$  &- & - & - & - & -0.534 & [-0.609, -0.468] & -0.631 & [-0.679, -0.583] \\
 $a_{2, 2}$  &- & - & - & - & 0.980 & [0.825, 1.286] & 0.942 & [0.817, 1.120] \\
 $a_{1, 1}^2\phi_{w1}$  & -  & - & - & - & 3.984 & [1.673, 7.448] & 4.203 & [2.972, 5.825] \\
 $a_{2, 2}^2\phi_{w2}$  & -  & - & - & - & 18.35 & [14.48, 23.31] & 19.43 & [16.63, 22.68] \\
 $\sigma_{\eta}^2\phi_{\eta}$ & -  & - & 0.079 & [0.030, 0.181] & - & - & 0.220 & [0.101, 0.444] \\
 $\tau_{1}^2$ & 0.784 & [0.663, 0.935] & 0.365 & [0.313, 0.426] & 0.073 & [0.022, 0.143] & 0.046 & [0.023, 0.069]\\
 $\tau_{2}^2$ & 0.920 & [0.778, 1.086] & 0.269 & [0.232, 0.311] & 0.023 & [0.008, 0.056] & 0.014 & [0.006, 0.027]\\
\hline
\hline
\end{tabular}
}
\end{center}
\label{tab:MT_est_S1=S2_all}
\end{table}

\begin{table}[htbp]
\caption{The estimation results for the simulated data with the holdout (I) in the $\mathcal{S}_{1}\neq \mathcal{S}_{2}$ case for low ($\sigma^2=1/3$) and high ($\sigma^2=1$) variance.}
\begin{center}
\scalebox{0.75}{
\begin{tabular}{lccccccccccc}
\hline
\hline
low &  & \multicolumn{2}{c}{$M_{1}^{=}$} & \multicolumn{2}{c}{$M_{2}^{=}$} & \multicolumn{2}{c}{$M_{3}^{=}$} & \multicolumn{2}{c}{$M_{4}^{=}$} \\
   & true & Mean &  $95\%$ Int  &  Mean &  $95\%$ Int & Mean &  $95\%$ Int  & Mean &  $95\%$ Int \\
\hline
 $\alpha_{1}$ & 6 & -  & - & 6.042 & [5.340, 6.490] & - & - & 5.409 & [3.050, 6.886] \\
 $\alpha_{2}$ & 1 & -  & - & 0.676 & [-0.650, 1.630] & - & - & 1.258 & [-0.381, 3.759] \\
 $\beta_{1, 1}$ & 0 & 0.213  & [0.144, 0.275] & 0.038 & [-0.637, 0.490] & 0.141 & [-0.793, 0.809] & -0.174 & [-2.393, 1.267] \\
 $\beta_{1, 2}$ & 0.5 & 0.711 & [0.488, 0.945] & 0.106 & [-1.180, 0.987] & 0.603 & [-0.407, 1.790] & 0.736 & [-0.641, 2.635] \\
 $\beta_{2, 1}$ & 0 & 0.036 & [0.004, 0.069] & -0.017 & [-0.206, 0.122] & 0.013 & [-0.284, 0.210] & -0.125 & [-1.008, 0.487] \\
 $\beta_{2, 2}$ & 0.5 & 0.638 & [0.518, 0.750] & 0.443 & [0.030, 0.740] & 0.600 & [0.263, 0.981] & 0.667 & [0.040, 1.585] \\
 $\gamma_{1}$ & 1 & - & - & 0.983 & [0.767, 1.268] & - & - & 0.933 & [0.613, 1.441] \\
 $\gamma_{2}$ & 0.3 & - & - & 0.314 & [0.223, 0.431] & - & - & 0.372 & [0.215, 0.608] \\
 $a_{1, 1}$ & - &- & - & - & - & 0.026 & [0.002, 0.154] & 0.001 & [0.000, 0.005] \\
 $a_{2, 1}$ & - &- & - & - & - & 0.608 & [0.353, 1.210] & 0.000 & [-0.008, 0.007] \\
 $a_{2, 2}$ & - &- & - & - & - &  0.188 & [0.110, 0.366] & 0.255 & [0.001, 0.897] \\
 $\sigma_{w1}^2\phi_{w1}$ & - & -  & - & - & - & 0.000 & [0.000, 0.003] & 0.000 & [0.000, 0.000] \\
 $\sigma_{w2}^2\phi_{w2}$ & - & -  & - & - & - & 0.072 & [0.030, 0.145] & 0.002 & [0.000, 0.024] \\
 $\sigma_{\eta}^2\phi_{\eta}$ & 1 &-  & - & 0.825 & [0.396, 1.573] & - & - & 0.658 & [0.171, 1.477] \\
 $\tau_{1}^2$ & 0.3 & 0.475  & [0.418, 0.534] & 0.295 & [0.257, 0.338] & 0.299 & [0.256, 0.346] & 0.297 & [0.242, 0.359]\\
 $\tau_{2}^2$ & 0.1 & 0.119  & [0.105, 0.135] & 0.100 & [0.087, 0.114] & 0.100 & [0.087, 0.113] & 0.103 & [0.085, 0.125]\\
\hline
\hline
high &  &\multicolumn{2}{c}{$M_{1}^{=}$} &  \multicolumn{2}{c}{$M_{2}^{=}$} & \multicolumn{2}{c}{$M_{3}^{=}$} & \multicolumn{2}{c}{$M_{4}^{=}$} \\
   & true & Mean &  $95\%$ Int  & Mean &  $95\%$ Int & Mean &  $95\%$ Int & Mean &  $95\%$ Int  \\
\hline
 $\alpha_{1}$ & 6 & -  & - & 5.469 & [4.917, 6.010] & - & - & 5.610 & [4.833, 6.153]\\
 $\alpha_{2}$ & 1 & -  & - & 1.654 & [0.610, 3.280] & - & - & 0.793 & [-0.429, 3.433] \\
 $\beta_{1, 0}$ & 0 & 0.443 & [0.359, 0.532] & -0.356 & [-1.015, 0.231] & -0.141 & [-1.395, 0.763] & 3.522 &[-1.053, 9.662] \\
 $\beta_{1, 1}$ & 0.5 & 0.761 & [0.448, 1.066] & 0.846 & [-0.252, 2.429] & -1.048 & [-3.676, 1.193] & 0.038 & [-1.391, 2.702]\\
 $\beta_{2, 0}$ & 0 & 0.125 & [0.089, 0.162] & -0.096 & [-0.279, 0.070] & -0.023 & [-0.394, 0.158] & -0.040 & [-0.320, 0.161] \\
 $\beta_{2, 1}$ & 0.5 & 0.579 & [0.461, 0.700] & 0.601 & [0.278, 1.071] & 0.058 & [-0.701, 0.729] & 0.338 & [-0.116, 1.203]\\
 $\gamma_{1}$ & 1 & -  & - & 1.104 & [0.913, 1.313] & - & - & 1.080 & [0.853, 1.359] \\
 $\gamma_{2}$ & 0.3 & - & - & 0.305 & [0.244, 0.374] & - & - & 0.320 & [0.232, 0.416]\\
 $a_{1, 1}$ & - & -  & - & - & - & 0.371 & [0.174, 1.064] & 9.373 & [0.124, 26.99] \\
 $a_{2, 1}$ & - & -  & - & - & - & 1.222 & [0.514, 2.935] & 0.324 & [-6.345, 6.828] \\
 $a_{2, 2}$ & - & -  & -  & - & - & 0.356 & [0.166, 0.842] & 0.036 & [0.000, 0.088] \\
 $\sigma_{w1}^2\phi_{w1}$ & - & -  & - & - & - & 0.974 & [0.045, 4.741] & 0.008 & [0.000, 0.061] \\
 $\sigma_{w2}^2\phi_{w2}$ & - & -  & - & - & - & 0.229 & [0.136, 0.385] & 0.000 & [0.000, 0.000] \\
 $\sigma_{\eta}^2\phi_{\eta}$ & 3 & -  & - & 2.428 & [1.655, 3.306] & - & - & 2.435 & [1.387, 4.091] \\
 $\tau_{1}^2$ & 0.3 & 0.812 & [0.711, 0.937] & 0.324 & [0.268, 0.391] & 0.309 & [0.238, 0.387] & 0.337 & [0.267, 0.416] \\
 $\tau_{2}^2$ & 0.1 & 0.125 & [0.109, 0.143] & 0.087 & [0.076, 0.100] & 0.085 & [0.073, 0.098] & 0.094 & [0.078, 0.113] \\
\hline
\hline
\end{tabular}
}
\end{center}
\label{tab:sim_est_S1neqS2}
\end{table}

\begin{table}[ht]
\caption{The estimation results for the MDBH-TPH data with the holdout (I) with $p=0.2$ in the $\mathcal{S}_{1}\neq \mathcal{S}_{2}$ case.}
\begin{center}
\scalebox{0.75}{
\begin{tabular}{lcccccccccc}
\hline
\hline
 & \multicolumn{2}{c}{$M_{1}^{=}$} & \multicolumn{2}{c}{$M_{2}^{=}$} & \multicolumn{2}{c}{$M_{3}^{=}$} & \multicolumn{2}{c}{$M_{4}^{=}$} \\
    & Mean &  $95\%$ Int  &  Mean &  $95\%$ Int & Mean &  $95\%$ Int  & Mean &  $95\%$ Int \\
\hline
 $\alpha_{0}$  & -  & - & 5.957 & [5.771, 6.158] & - & - & 5.725 & [4.241, 7.926] \\
 $\alpha_{1}$  & -  & - & -1.557 & [-5.273, 0.703] & - & - & 3.218 & [-0.639, 6.522] \\
 $\alpha_{2}$  & -  & - & -1.724 & [-4.473, 1.610] & - & - & -0.132 & [-3.167, 2.540] \\
 $\alpha_{3}$  & -  & - & -0.415 & [-2.986, 2.080] & - & - & 1.490 & [-1.621, 3.536] \\
 $\alpha_{4}$  & -  & - & -1.280 & [-5.213, 1.595] & - & - & -2.972 & [-6.369, 0.232] \\
 $\beta_{1, 0}$  & 0.855 & [0.428, 1.289] & -0.572 & [-1.453, 0.523] & 1.234 & [0.532, 1.896] & -0.633 & [-2.071, 1.612] \\
 $\beta_{1, 1}$  & 15.73 & [5.033, 25.62] & 17.98 & [4.560, 30.37] & 4.000 & [-2.699, 10.82] & 17.74 & [3.390, 30.67] \\
 $\beta_{1, 2}$  & -0.792 & [-2.326, 0.743] & 0.833 & [-1.521, 3.017] & 1.034 & [-0.159, 2.159] & 1.011 & [-1.052, 3.093] \\
 $\beta_{1, 3}$  & 5.469 & [3.382, 7.475] & 10.62 & [7.721, 13.33] & 4.229 & [2.441, 5.824] & 9.690 & [6.606, 12.76] \\
 $\beta_{1, 4}$  & -1.379 & [-2.806, 0.101] & -0.597 & [-2.993, 1.581] & -0.977 & [-2.102, 5.824] & -1.269 & [-3.317, 0.937] \\
 $\beta_{2, 0}$  & 8.062 & [7.349, 8.764] & -0.221 & [-1.573, 0.879] & 8.307 & [7.650, 8.989] & 0.082 & [-0.889, 0.961] \\
 $\beta_{2, 1}$  & -1.483 & [-17.93, 15.50] & -0.650 & [-15.25, 11.72] & -7.861 & [-21.26, 4.683] & -4.779 & [-16.95, 8.418] \\
 $\beta_{2, 2}$  & -0.805 & [-4.533, 3.118] & -2.816 & [-5.049, -0.652] & -2.409 & [-4.524, 0.162] & -2.194 & [-4.350, -0.077] \\
 $\beta_{2, 3}$  & -4.540 & [-8.819, -0.129] & -0.672 & [-3.742, 2.364] & -2.840 & [-6.294, 0.648] & -2.352 & [-5.235, 0.632] \\
 $\beta_{2, 4}$  & 4.082 & [0.210, 7.726] & 0.683 & [-1.475, 2.909] & 1.795 & [-0.681, 4.317] & 0.333 & [-1.660, 2.094] \\
 $\gamma_{1}$  & - & - & 9.178 & [5.150, 14.50] & - & - & 0.394 & [-0.490, 1.129] \\
 $\gamma_{2}$  & - & - & -11.68 & [-18.14, -6.571] & - & - & 0.113 & [-0.522, 1.199] \\
 $a_{1, 1}$  &- & - & - & - & 0.544 & [0.213, 1.456] & 1.331 & [0.421, 4.192] \\
 $a_{2, 1}$  &- & - & - & - & -0.423 & [-0.517, -0.354] & -1.108 & [-1.254, -0.966] \\
 $a_{2, 2}$  &- & - & - & - & 1.206 & [1.039, 1.424] & 0.978 & [0.842, 1.176] \\
 $a_{1, 1}^2\phi_{w1}$  & -  & - & - & - & 0.502 & [0.242, 0.892] & 3.592 & [1.353, 6.813] \\
 $a_{2, 2}^2\phi_{w2}$  & -  & - & - & - & 26.84 & [21.67, 32.55] & 18.06 & [14.33, 22.67] \\
 $\sigma_{\eta}^2\phi_{\eta}$ & -  & - & 0.075 & [0.020, 0.237] & - & - & 0.368 & [0.067, 0.731] \\
 $\tau_{1}^2$ & 0.175 & [0.146, 0.208] & 0.349 & [0.282, 0.430] & 0.018 & [0.009, 0.027] & 0.078 & [0.023, 0.147]\\
 $\tau_{2}^2$ & 1.393 & [1.170, 1.649] & 0.227 & [0.182, 0.280] & 0.007 & [0.001, 0.025] & 0.024 & [0.008, 0.055]\\
\hline
\hline
\end{tabular}
}
\end{center}
\label{tab:MT_est_S1neqS2_random}
\end{table}

\begin{table}[htbp]
\caption{The estimation results for the simulated data in the disjoint $\mathcal{S}_{1}\neq \mathcal{S}_{2}$ case for low ($\sigma^2=1/3$) and high ($\sigma^2=1$) variance.}
\begin{center}
\scalebox{0.75}{
\begin{tabular}{lccccccccccc}
\hline
\hline
low &  & \multicolumn{2}{c}{$M_{1}^{*\neq}$} & \multicolumn{2}{c}{$M_{2}^{*\neq}$} \\
   & true & Mean &  $95\%$ Int  & Mean &  $95\%$ Int \\
\hline
 $\alpha_{1, 1}$ & 6 & - & - & 5.965 & [5.330, 7.132] \\
 $\alpha_{2, 1}$ & 1 & - & - & 1.763 & [0.295, 4.035] \\
 $\alpha_{1, 2}$ & 6 & - & - & 6.109 & [3.068, 7.957] \\
 $\alpha_{2, 2}$ & 1 & - & - & 2.591 & [-0.154, 5.098] \\
 $\beta_{1, 1}$ & 0 & 0.141 & [-0.793, 0.809] & 0.872 & [-0.749, 5.340] \\
 $\beta_{1, 2}$ & 0.5 & 0.603 & [-0.407, 1.790] & 1.427 & [0.129, 3.404] \\
 $\beta_{2, 1}$ & 0 & 0.013 & [-0.284, 0.210] & 0.371 & [-0.227, 0.773] \\
 $\beta_{2, 2}$ & 0.5 & 0.600 & [0.263, 0.981] & 1.040 & [0.205, 1.797] \\
 $\gamma_{1}$ & 1 & - & - & 0.958 & [0.703, 1.270] \\
 $\gamma_{2}$ & 0.3 & - & - & 0.301 & [0.243, 0.372] \\
 $a_{1, 1}$ & - & 0.026 & [0.002, 0.154] & 0.171 & [0.017, 0.436] \\
 $a_{2, 1}$ & - & 0.608 & [0.353, 1.210] & 0.038 & [-1.966, 1.736] \\
 $a_{2, 2}$ & - & 0.188 & [0.110, 0.366] & 0.676 & [0.005, 1.383] \\
 $\sigma_{w1}^2\phi_{w1}$ & - & 0.000 & [0.000, 0.003] & 0.022 & [0.000, 0.319] \\
 $\sigma_{w2}^2\phi_{w2}$ & - & 0.072 & [0.030, 0.145] & 0.000 & [0.000, 0.116] \\
 $\sigma_{\eta_{1}}^2\phi_{\eta_{1}}$ & 1 & - & - & 0.992 & [0.449, 1.928] \\
 $\sigma_{\eta_{2}}^2\phi_{\eta_{2}}$ & 1 & - & - & 1.477 & [0.755, 2.416] \\
 $\tau_{1}^2$ & 0.3 & 0.299 & [0.256, 0.346] & 0.305 & [0.262, 0.359]\\
 $\tau_{2}^2$ & 0.1 & 0.100 & [0.087, 0.113] & 0.093 & [0.081, 0.107]\\
\hline
\hline
high &  & \multicolumn{2}{c}{$M_{1}^{*\neq}$} & \multicolumn{2}{c}{$M_{2}^{*\neq}$} \\
   & true & Mean &  $95\%$ Int & Mean &  $95\%$ Int  \\
\hline
 $\alpha_{1, 1}$ & 6 & - & - & 6.180 & [5.094, 7.512]\\
 $\alpha_{2, 1}$ & 1 & - & - & -0.343 & [-1.694, 1.991] \\
 $\alpha_{1, 2}$ & 6 & - & - & 6.022 & [4.987, 6.545]\\
 $\alpha_{2, 2}$ & 1 & - & - & 1.508 & [-0.077, 2.772] \\
$\beta_{1, 0}$ & 0 & -0.141 & [-1.395, 0.763] & 0.169 & [-1.583, 1.732] \\
 $\beta_{1, 1}$ & 0.5 & -1.048 & [-3.676, 1.193] & -0.411 & [-1.876, 2.492]\\
 $\beta_{2, 0}$ & 0 & -0.023 & [-0.394, 0.158] & -0.016 & [-0.376, 0.215] \\
 $\beta_{2, 1}$ & 0.5 & 0.058 & [-0.701, 0.729] & 0.542 & [-0.061, 0.954]\\
 $\gamma_{1}$ & 1 & - & - & 0.996 & [0.763, 1.216] \\
 $\gamma_{2}$ & 0.3 & - & - & 0.331 & [0.241, 0.439]\\
 $a_{1, 1}$ & - & 0.371 & [0.174, 1.064] & 0.040 & [0.000, 0.117] \\
 $a_{2, 1}$ & - & 1.222 & [0.514, 2.935] & 0.020 & [-0.014, 0.060] \\
 $a_{2, 2}$ & - & 0.356 & [0.166, 0.842] & 0.133 & [0.041, 0.245] \\
 $\sigma_{w1}^2\phi_{w1}$ & - & 0.974 & [0.045, 4.741] & 0.000 & [0.000, 0.000] \\
 $\sigma_{w2}^2\phi_{w2}$ & - & 0.229 & [0.136, 0.385] & 0.019 & [0.001, 0.111] \\
 $\sigma_{\eta_{1}}^2\phi_{\eta_{1}}$ & 3 & - & - & 2.286 & [1.279, 3.469] \\
 $\sigma_{\eta_{2}}^2\phi_{\eta_{2}}$ & 3 & - & - & 2.038 & [0.856, 3.345] \\
 $\tau_{1}^2$ & 0.3 & 0.309 & [0.238, 0.387] & 0.319 & [0.271, 0.371] \\
 $\tau_{2}^2$ & 0.1 & 0.085 & [0.073, 0.098] & 0.097 & [0.084, 0.111] \\
\hline
\hline
\end{tabular}
}
\end{center}
\label{tab:sim_est_S1disS2}
\end{table}

\end{document}